\documentclass[aps,prd,twocolumn,showpacs,floats,nofootinbib,10pt,superscriptaddress]{revtex4-2}                		
\usepackage{graphicx}				
\usepackage{amssymb}
\usepackage{amsmath}
\usepackage{amsfonts,mathrsfs}
\usepackage{mathtools}
\usepackage{lipsum}
\usepackage{multirow}
\usepackage{indentfirst}
\usepackage{graphicx}
\usepackage{subcaption}
\graphicspath{{./Figures/}}
\usepackage[table]{xcolor}
\usepackage[
    colorlinks=true,
    linkcolor=blue
]{hyperref}
\usepackage[section]{placeins}
\usepackage{placeins}
\usepackage{float} 
\usepackage{caption,subcaption} 
\usepackage{orcidlink}

\newtheorem{RC}{Robustness Criterion}

\let\Oldsection\section
\renewcommand{\section}{\FloatBarrier\Oldsection}

\let\Oldsubsection\subsection
\renewcommand{\subsection}{\FloatBarrier\Oldsubsection}

\let\Oldsubsubsection\subsubsection
\renewcommand{\subsubsection}{\FloatBarrier\Oldsubsubsection}
\DeclarePairedDelimiter\abs{\lvert}{\rvert}
\DeclarePairedDelimiter\norm{\lVert}{\rVert}
\makeatletter
\let\oldabs\abs
\def\abs{\@ifstar{\oldabs}{\oldabs*}}

\let\oldnorm\norm
\def\norm{\@ifstar{\oldnorm}{\oldnorm*}}
\makeatother

\let\Re\relax
\DeclareMathOperator{\Re}{{Re}}
\let\Im\relax
\DeclareMathOperator{\Im}{{Im}}

\newcommand{\swY}[4][]{{}_{{}_{#2}}\!Y^{#1}_{#3}(#4)}
\newcommand{\YSH}[4][]{{}_{#2}\mathcal{A}^{#1}_{#3}(#4)}
\newcommand{\swSH}[5][]{{}_{{}_{#2}}S^{#1}_{#3}(#4;#5)}
\newcommand{\YSHn}[2][]{\mathcal{A}^{#1}_{#2}}

\usepackage{xcolor}

\begin{document}

\title{Robustness of extracting quasinormal mode information from black hole merger simulations}

\author{Leda Gao\orcidlink{0000-0003-3319-166X}}
\email{gaol18@wfu.edu}
\author{Gregory B. Cook\orcidlink{0000-0002-4395-7617}}
\email{cookgb@wfu.edu}
\affiliation{Department of Physics, Wake Forest University, Winston-Salem, North Carolina 27109, USA}
\author{Lawrence E. Kidder\orcidlink{0000-0001-5392-7342}}
\affiliation{Cornell Center for Astrophysics and Planetary Science, Cornell University, Ithaca, New York 14853, USA}
\author{Harald P. Pfeiffer\orcidlink{0000-0001-9288-519X}}
\affiliation{Max Planck Institute for Gravitational Physics (Albert Einstein Institute), Am Muhlenberg 1, Potsdam 14476, Germany}
\author{Mark A. Scheel\orcidlink{0000-0001-6656-9134}}
\affiliation{Theoretical Astrophysics 350-17, California Institute of Technology, Pasadena, California 91125, USA}
\author{Nils Deppe\orcidlink{0000-0003-4557-4115}}
\affiliation{Department of Physics, Cornell University, Ithaca,
  New York 14853, USA}
\author{William Throwe\orcidlink{0000-0001-5059-4378}}
\affiliation{Cornell Center for Astrophysics and Planetary Science, Cornell University, Ithaca, New York 14853, USA}
\author{Nils L. Vu\orcidlink{0000-0002-5767-3949}}
\affiliation{Theoretical Astrophysics 350-17, California Institute of Technology, Pasadena, California 91125, USA}
\author{Kyle C. Nelli\orcidlink{0000-0003-2426-8768}}
\affiliation{Theoretical Astrophysics 350-17, California Institute of Technology, Pasadena, California 91125, USA}
\author{Jordan Moxon\orcidlink{0000-0001-9891-8677}}
\affiliation{Theoretical Astrophysics 350-17, California Institute of Technology, Pasadena, California 91125, USA}
\author{Michael Boyle\orcidlink{0000-0002-5075-5116}}
\affiliation{Cornell Center for Astrophysics and Planetary Science, Cornell University, Ithaca, New York 14853, USA}

\begin{abstract}
In linear perturbation theory, the ringdown of a gravitational wave (GW) signal is described by a linear combination of quasinormal modes (QNMs). Detecting QNMs from GW signals is a promising way to test GR, central to the developing field of black-hole spectroscopy. More robust black hole spectroscopy tests could also consider the ringdown amplitude-phase consistency.  That requires an accurate understanding of the excitation and stability of the QNM expansion coefficients. In this paper, we investigate the robustness of the extracted $m=2$ QNM coefficients obtained from a high-accuracy numerical relativity waveform.  We explore a framework to assess the robustness of QNM coefficients.  Within this framework, we not only consider the traditional criterion related to the constancy of a QNM's expansion coefficients over a window in time, but also emphasize the importance of consistency among fitting models. In addition, we implement an iterative greedy approach within which we fix certain QNM coefficients.  We apply this approach to linear fitting, and to nonlinear fitting where the properties of the remnant black hole are treated as unknown variables.  We find that the robustness of overtone coefficients is enhanced by our greedy approach, particularly for the $(2,2,2,+)$ overtone. Based on our robustness criteria applied to the $m=2$ signal modes, we find the $(2\!\sim\!4,2,0,+)$ and $(2,2,1\!\sim\!2,+)$ modes are robust, while the $(3,2,1,+)$ subdominant mode is only marginally robust. After we subtract the contributions of the $(2\!\sim\!4,2,0,+)$ and $(2\!\sim\!3,2,1,+)$ QNMs from signal mode $(4,2)$, we also find evidence for the quadratic QNM $(2,1,0,+)\times(2,1,0,+)$.
\end{abstract}

\maketitle

\section{Introduction} \label{Sec: Intro}
Binary black hole coalescence is one major source of gravitational wave signals for current and next-generation gravitational wave detectors. After the binary black holes merge, this dynamical system settles down to a remnant Kerr black hole.  The process of settling down to the remnant black hole (BH) is referred to as the ringdown phase, and it starts at a time near the peak intensity of the gravitational wave (GW) signal emitted during the binary black hole (BBH) merger.  After nonlinear effects from the BBH merger become subdominant, and before the onset of power-law tails, there should be a period of time during the ringdown signal that can be described accurately by linear BH perturbation theory\cite{Teukolsky:1972my,Teukolsky:1973,Press:1973zz,Leaver:1986gd}.  During this period of time, the GW signal emitted by the perturbed remnant Kerr BH is modeled by a linear combination of quasinormal modes (QNMs), with the complex QNM frequencies $\omega_{\ell m n}$ only depending on the mass and angular momentum of the remnant BH.  The latter property embodies the BH no-hair theorem.  When we use a linear superposition of QNMs to describe the GW ringdown waveform, each QNM has an associated QNM expansion coefficient, $C_{\ell m n}$, which is complex and describes the amplitude $A_{\ell m n}$ and phase $\phi_{\ell m n}$ of the mode.  The relative excitation of the different QNM amplitudes and phases depend on the astrophysical process leading to the ringdown.  This means that measuring the relative excitation of these QNMs should encode information about the properties of the progenitor system, the binary system prior to the ringdown\cite{Kamaretsos:2011um,Kamaretsos:2012bs,London:2014cma}.  An accurate and reliable means to fully analyze GW ringdown signals can serve as a powerful tool to test general relativity and understand the nature of compact binary mergers. This idea has been referred to as black hole spectroscopy\cite{Dreyer:2003bv}. 

Many approaches have been explored for fitting ringdown signals.  One general approach assumes the remnant BH is Kerr, in which case it is possible to fit for the QNM expansion coefficients either assuming the mass and angular momentum of the BH are known, or to fit for these remnant properties as well~\cite{Giesler:2019uxc,Cook:2020otn,MaganaZertuche:2021syq}.  Alternatively, it is possible to simply fit for a set of damped sinusoids which lead to methods which are agnostic as to the correct theory of gravity~\cite{Baibhav:2023clw,Cheung:2023vki,Giesler:2024hcr}.  Both general approaches can be applied to fitting both experimental GW data and to numerically simulated waveforms.  In this paper, we explore the case of fitting numerically generated ringdown waveforms under the assumption that general relativity is the correct theory of gravity.  Such fitting is of interest because catalogs of simulated waveforms have been generated, with the contained datasets spanning ever-increasing portions of the parameter space of BBH systems.  Determining the QNM expansion coefficients for each dataset's ringdown signal is a first step in constructing surrogate ringdown waveforms parametrized by the properties of the progenitor BBH system~\cite{Babak:2016tgq,Varma:2018mmi,MaganaZertuche:2024ajz,Pacilio:2024tdl}.  Such surrogate waveforms are then useful in estimating the properties of the progenitor BBH system associated with an experimentally detected GW ringdown signal and for testing the no-hair theorem.

There have already been several efforts made to measure, in the observational ringdown signals, QNMs beyond the most dominant mode.  These studies have considered both overtones of the dominant mode and subdominant QNMs\cite{Isi:2019aib,Cotesta:2022pci,Finch:2022ynt,Isi:2022mhy,Ma:2023vvr,Capano:2021etf,Capano:2022zqm}. However, due to the sensitivity limits of the current generation of GW detectors, there are very few events that show even tentative evidence for the detection of the first overtone or subdominant QNMs.  While the traditional way to test GR with BH spectroscopy has focused on checking the consistency of measured QNM frequencies to the predicted frequencies from BH perturbation theory, more robust tests could also consider the ringdown amplitude-phase consistency\cite{Forteza:2022tgq}. In such a test, the consistency of measured QNM amplitudes and phases from observation are checked against the predictions made from GR. To implement this consistency test, we need a thorough understanding of the excitation of each QNM's amplitude and phase for different binary systems under the assumption that GR is the correct theory of gravity.  While QNM frequencies, and their associated angular functions, for Kerr black holes can now be calculated with very high precision by solving the Teukolsky equation\cite{Cook:2014cta}, computation of the excitation of each mode from first principles is much more difficult.  At present the main method to understand the excitation of the QNM amplitudes and phases is to fit the ringdown portion of numerical relativity (NR) simulated waveforms with a superposition of QNMs.

In linear perturbation theory, the QNM expansion coefficients $C_{\ell m n}$ of a ringdown signal have no time dependence.  If we fit for the expansion coefficients using a portion of the signal that covers only the latter portion of the ringdown, then nonlinear effects will be minimized, but simulation errors will become more significant and only the slowest decaying modes will be present.  If we include earlier portions of the ringdown signal in the fitting process, then the more rapidly decaying modes may be present, but it is likely that nonlinear effects will also be more significant within the signal.  Numerical errors and the presence of nonlinear effects will cause all attempts at fitting to yield QNM expansion coefficients with values that depend on the range of ringdown signal used for each fit\cite{Baibhav:2023clw}.  A common practice is to fit using a range of data $t_i\le t\le t_e$, and to consider a range of start times $t_i$ and a common end time $t_e$.  The fit expansion coefficients can then be considered as functions of the fit-start time $C_{\ell m n}(t_i)$, and it is necessary to determine which fit values are faithful to the actual linear ringdown signal.  We will refer to such a QNM which can be faithfully fitted as ``robust''.  Since linear perturbation theory predicts QNM expansion coefficients which are constants, we should expect reasonable fit coefficients to be nearly constant as a function of the fit-start time $t_i$.  The most common method used to decide whether a given QNM expansion coefficient is robust or not is to evaluate the stability of each fitted QNM coefficients throughout an extraction window $t_-\le t_i\le t_+$\cite{Baibhav:2023clw,Cheung:2023vki,Zhu:2023fnf,Zhu:2024rej}.  However, it seems likely that this criteria is a necessary but not sufficient criteria for accurately extracting QNM expansion coefficient\cite{Zhu:2024rej}. The primary goal of this paper is to explore how we might determine if a given QNM is robust.

To better understand the uncertainty of the extracted QNM coefficients, in addition to the use of extraction windows, we also consider the uncertainty among extracted coefficients $C_{\ell m n}$ from different fitting models and different fitting methods.  We define a fitting model as a specific combination of a set of signal modes and a set of fitting QNMs.  The NR simulated waveform is decomposed into spin-weight $-2$ spherical harmonic modes $C_{\ell m}(t)$, while the QNMs are functions of the spin-weight $-2$ spheroidal harmonics.  As a result, a single QNM in the fitting function will contribute to multiple signal modes, an effect known as spherical-spheroidal mode mixing, and it is useful to simultaneously fit a given fitting function to multiple source modes\cite{Cook:2020otn,Dhani:2021vac,MaganaZertuche:2021syq}.  So far, work in this area has primarily focused on how multimode fitting can increase the overlap (lower the mismatch) between the signal and the fitting function, and improve the accuracy of the estimated remnant parameters.  More recent studies investigating the stability of the extracted QNM coefficients\cite{Baibhav:2023clw,Cheung:2023vki,Takahashi:2023tkb,Clarke:2024lwi} take only part of the mode-mixing effect into account.  The notion of multimode fitting as fitting multiple signal modes simultaneously to the same set of QNMs was first explore in Ref.~\cite{Cook:2020otn}.  As far as we know, the effect of this notion of multimode fitting on the robustness of the extracted QNM coefficients has not been fully investigated.  We find that the robustness of the QNM coefficients, especially the subdominant fundamental modes, can be improved after we fully consider the mode-mixing effect. 

Because the QNMs are neither orthogonal nor complete, the choice of the set of QNMs to include in the fitting function also affects the fit results.  Although algorithms have been developed to automatically pick the combinations of modes based on different goals\cite{MaganaZertuche:2021syq,Cheung:2023vki}, it seems that choosing the modes for fitting is a nontrivial task.  Since we do not know \textit{a priori} which QNMs are physically contributing to specific time segments of the ringdown signal, there are always some extra QNMs included or some effects unconsidered for different models. While the extra QNMs tend to be overfitted, unconsidered effects might be fit away by the QNMs included during the fitting. This ambiguity in choosing the QNM fitting model is almost unavoidable. Therefore, in this paper, we compare the extracted QNM coefficients obtained from different fitting models systematically.

Comparing the results from different fitting models requires that we generate some statistics associated with each model.  This is accomplished through fixing an extraction window for each mode or set of modes.  Multiple fits, each with a different start time $t_i$, are performed within this extraction window and this provides one notion of a set of data which yields a distribution of fit results.  A common approach is to use some measure of the constancy of a QNM's expansion coefficient to choose its extraction window.  As an alternative, we will also explore the use of nonlinear fitting, which also extracts the remnant parameters, to provide a different definition of the extraction window.
 
We will also explore different fitting methods.  Our methods are based on the ``eigenvalue method'' described in Ref.~\cite{Cook:2020otn}.  It is very similar to standard linear least squares fitting, but more correctly handles the angular behavior of the QNMs.  When the remnant's parameters are taken as known, the eigenvalue method reduces to solving a linear problem and we refer to it as linear fitting.  With standard linear fitting, all of the modes in a given fitting function are determined simultaneously.  An alternative approach is to iteratively extract the QNM expansion coefficient from the ringdown signal. Ma \textit{et~al}. developed frequency-domain filters to remove corresponding QNMs from the original ringdown signal\cite{Ma:2022wpv,Ma:2023cwe,Ma:2023vvr}. Their methods focus more on investigating BH spectroscopy based on the QNM frequencies. Takahashi \textit{et~al}.\cite{Takahashi:2023tkb} implemented an iterative procedure consisting of fitting and subtraction off the longest-lived mode of the ringdown waveform in the time domain.  In our work, we implement a similar greedy approach in which we fix certain QNM expansion coefficients during the fitting.  This has the benefit of lowering the degrees of freedom for a given fitting problem, and can be easily implemented within the eigenvalue method.  We refer to this fitting method as linear greedy.  Both methods can be generalized to nonlinear fitting where the mass and angular momentum of the black hole are also extracted by maximizing, over the remnant parameter space, the overlap between the signal and the fitting function.  We refer to the nonlinear variants of both linear methods simply as nonlinear fitting and nonlinear-greedy fitting.

We find that using the greedy approach to fix the lowest overtones first, and then progressing to successively higher overtones, is beneficial.  This approach is motivated by two aspects of the QNMs.  First, the QNMs with the lowest overtones typically have the longest decay times and are the most robust modes.  The second aspect is a property of the spin-weighted spheroidal harmonics, proven by London\cite{London:2020uva}, that the spheroidal harmonics for QNMs with the same overtone value form a minimal set.  On the other hand, modes with the same $\ell$ and $m$ but different overtone values are not minimal, which means that they do not encode distinct mode information. Using this approach, we find that the robustness of the higher overtones' coefficients is improved. But, for the NR dataset we explore, the last mode we can extract with marginal robustness is the prograde QNM $(2,2,2)$.  Any higher overtones can only be extracted with large uncertainty.  That matches the conclusion drawn from several recent papers\cite{Cheung:2023vki,Zhu:2023mzv,Clarke:2024lwi}. On the other hand, a recent paper~\cite{Giesler:2024hcr} states that the difficulty in finding stable higher overtones is due to the ill-conditioned nature of fitting damped exponentials in the presence of even small amounts of noise. More discussion about this work is in the conclusion section. In Appendix~\ref{app:pure_damped_waveforms}, we include supplemental tests of the greedy approach by applying it to model waveforms consisting of linear combinations of QNMs with known amplitudes and phases. The benefits of the greedy approach are further confirmed in this controlled setting. 

The outline of this paper is as follows. Section~\ref{sec: Method} presents the details about the waveform fitting method and how we implement the greedy algorithm, including the conventions and definitions. In Sec.~\ref{sec: Linear_Fitting}, we apply the linear fitting method to the simulation waveform SXS:BBH\_ExtCCE:0305. We also present our definition and criteria for the robustness of the QNM coefficients. A thorough investigation about multimode fitting and the implementation of the greedy algorithm is also presented in this section. In Sec.~\ref{sec:nonlinear_fitting}, we generalized the fitting method to the nonlinear fitting scenario and present the results obtained from nonlinear fitting with the implementation of the greedy algorithm. Section~\ref{sec: conclusion} presents the conclusion of the results described in the previous section. In Appendix~\ref{app:details_rescaling}, we describe the details of a rescaling method implemented during the waveform fitting, which can lower the numerical noise of the fitting results. In Appendix~\ref{app:pure_damped_waveforms}, we performed linear fitting to several model waveforms and compared the performance of regular linear fitting with the linear greedy approach.

\section{Method}\label{sec: Method}
\subsection{Convention and definitions}
In this work, we focus on fitting gravitational-wave ringdown signals simulated by numerical relativity. We follow the conventions detailed in paper~\cite{Cook:2020otn}.

The gravitational-wave information can be presented in terms of several different quantities, such as the gravitational strain $h$, the Newman-Penrose scaler $\Psi_4$, and the Bondi news function $\mathcal{N}$. The simulated waveforms used in our work are provided in terms of spin-weight $-2$ spherical harmonic modes 
\begin{equation}\label{eqn:psiNR}
\psi_{\rm NR} = \sum_{\ell{m}}{C_{\ell{m}}(t)\,\swY{-2}{\ell{m}}{\theta,\phi}}.
\end{equation}
We will refer to the harmonic modes $C_{\ell{m}}$ of the numerical relativity signal as signal modes. 
Although we focus solely on gravitational strain $h$, we can apply our fitting method to other quantities that represent gravitational wave information.

In the regime of linear perturbation theory, when the waveform extraction coordinates align with the remnant black hole's spin, the simulated ringdown waveform can be expressed in terms of quasinormal modes(QNMs) as~\cite{Berti:2005ys}  
\begin{eqnarray}\label{eqn:QNMexpansion1}
\psi_{\text{fit}} &=& \sum_{\ell{m}n}\Bigl\{C^+_{\ell{m}n}e^{-i\omega^+_{\ell{m}n} (t-r^*)}
    \swSH{-2}{\ell{m}}{\theta^\prime,\phi^\prime}{a\omega^+_{\ell{m}n}} \nonumber\\
    &&
    + C^-_{\ell{m}n}e^{-i\omega^-_{\ell{m}n} (t-r^*)}
    \swSH{-2}{\ell{m}}{\theta^\prime,\phi^\prime}{a\omega^-_{\ell{m}n}}
    \Bigr\},
\end{eqnarray}
where the angular basis functions are the spin-weight $-2$ spheroidal harmonics $\swSH{-2}{\ell{m}}{\theta^\prime,\phi^\prime}{c}$ with indices $(\ell,m)$. Here, $\theta^\prime$ and $\phi^\prime$ are polar coordinates measured relative to the angular momentum of the remnant Kerr black hole. In $\swSH{-2}{\ell{m}}{\theta^\prime,\phi^\prime}{c}$, $c$ is the spheroidal parameter, which is $c=a\omega^{\pm}_{\ell{m}n}$ in this case. $a$ is the angular momentum parameter $a=J_f/M_f$, where $M_f$ and $J_f$ are the mass and angular momentum of the remnant black hole. 

The complex frequency $\omega_{\ell mn}$ associated with each QNM are split into two families of modes: ``ordinary'' modes $\omega^{+}_{\ell m n}$ and ``mirror'' modes $\omega^{-}_{\ell m n}$. The ordinary and mirror modes are related by
\begin{equation}\label{eqn: ordinary_mirror_modes}
\omega^+_{\ell mn}=-(\omega^-_{\ell (-m)n})^* \equiv \omega_{\ell mn}. 
\end{equation}
Note the definition of the mode frequency $\omega^+_{\ell mn}$ without a superscript sign as representing the ordinary family of modes.  This notation is used because QNM data is typically only stored for the $\omega^+_{\ell mn}$ modes and will be used in expressions that have been transformed to use only this family of modes.

The QNM frequencies are completely determined by the mass and spin of the remnant BH in the regime of GR. Some works also classify these QNM frequencies as prograde and retrograde modes based on the circulating direction of the QNMs' wave fronts\cite{MaganaZertuche:2021syq,Li:2021wgz,Cheung:2023vki} although there is still no universal agreement on defining the prograde and retrograde modes.  In this work, we adopt the convention established in paper\cite{MaganaZertuche:2021syq} for the classification of the prograde modes and retrograde modes. According to this convention, surfaces of constant phase for prograde modes corotate with respect to the rotation of the remnant BH, while the retrograde modes counter-rotate with respect to it. This means that $\omega^+_{\ell{m}n}$ modes with $m>0$, and their mirror modes $\omega^-_{\ell(-m)n}$, are defined as prograde modes.  Similarly, $\omega^+_{\ell{m}n}$ modes with $m<0$, and their mirror modes, are defined as retrograde modes.

Each QNM is associated with a complex QNM expansion coefficient, $C^{\pm}_{\ell{m}n}$. It is useful to express it as amplitude $A^{\pm}_{\ell mn}$ and phase $\phi^{\pm}_{\ell mn}$,
\begin{equation} \label{eqn: Clmn_amp_phase}
    C^{\pm}_{\ell mn}=A^{\pm}_{\ell mn}e^{i \phi^{\pm}_{\ell mn}}.
\end{equation}
The time-dependent part $e^{-i\omega^{\pm}_{\ell{m}n} (t-r^*)}$ of each QNM is a damped exponential depending on the QNM frequency, where $t-r^*$ is the retarded time expressed in terms of the tortoise-coordinate $r^*$.

Since the spin-weighed spherical harmonics form a complete basis, the spin-weighted spheroidal harmonics can be expanded in terms of the spin-weighted spherical harmonics~\cite{Teukolsky:1973,Cook:2014cta},
\begin{equation} \label{eqn:swSH:expan}
\swSH{-2}{\ell{m}}{\theta^\prime,\phi^\prime}{c} = \sum_{\acute\ell}
\YSH{-2}{\acute\ell\ell{m}}{c} \swY{-2}{\acute\ell{m}}{\theta^\prime,\phi^\prime}.
\end{equation}
The spheroidal-harmonic expansion coefficients $\YSH{-2}{\acute\ell\ell{m}}{c}$ are functions of the spheroidal parameter $c$. In some papers~\cite{Berti:2014fga}, this coefficient is also called the spherical-spheroidal mixing coefficient. 

The QNM datasets we use for $\omega^{+}_{\ell{m}n}$ and $\YSH{-2}{\acute\ell\ell{m}}{c}$ are publicly accessible through Zenodo\cite{Cook:2024zenodo}. These dataset only contain information for the ordinary QNM family.  Data for the mirror modes can be obtained using Eq.~\eqref{eqn: ordinary_mirror_modes} and the corresponding transformation for $\YSH{-2}{\acute\ell\ell{m}}{a\omega^-_{\ell mn}}$ is 
\begin{equation}\label{eqn: expansion_mirror}
    \YSH{-2}{\acute\ell\ell{m}}{a\omega^-_{\ell mn}}=(-1)^{\acute\ell+\ell}\YSH[*]{-2}{\acute\ell\ell(-m)}{a\omega^+_{\ell(-m)n}},
\end{equation}
using various symmetry properties\cite{Cook:2014cta,Cook:2020otn}. 
In the following sections, $\YSH{-2}{\acute\ell\ell{m}}{a\omega^+_{\ell mn}}$ is simplified to be $\YSHn{\acute{\ell}\ell m n}$.  Finally, we fix the phase freedom in the spin-weighted spheroidal harmonics using the method specified by Cook and Zalutskiy\cite{Cook:2014cta}, whereby the expansion coefficient $\YSHn{\ell\ell m n}$ is taken to be real and positive.  While this has been a standard choice, we note that it may not be be optimal choice\cite{cookwang:2025} for future work.

Beyond linear perturbation, it is possible to include quadratic QNMs, arising from second-order perturbations~\cite{London:2014cma,Mitman:2022qdl,Cheung:2022rbm}, within our fitting model. The frequency of quadratic QNM is the linear combination of its parent QNMs. The frequencies of quadratic QNMs, which we label as $(\ell_1,m_1,n_1,\pm)\times(\ell_2,m_2,n_2,\pm)$, can be expressed as 
\begin{equation}
\omega_{(\ell_1,m_1,n_1,\pm)\times(\ell_2,m_2,n_2,\pm)}=\omega^{\pm}_{\ell_1 m_1n_1}+\omega^{\pm}_{\ell_2 m_2n_2}.
\end{equation}
Correctly handling the angular behavior of these modes is beyond the scope of this work, and we simply take the spheroidal-harmonic expansion coefficients to be defined as $\delta_{\acute{\ell}(\ell_1+\ell_2)}$. Thus, in our fitting process, the quadratic QNM $(\ell_1,m_1,n_1,\pm)\times(\ell_2,m_2,n_2,\pm)$ only contributes to signal mode $C_{(\ell_1+\ell_2)(m_1+m_2)}$.  

To simplify the expressions in Eq.~\eqref{eqn:QNMexpansion1}, our fitting function will be written as 
\begin{equation}\label{eqn:psi_fit_k}
\psi_{\rm fit} = \sum_{k\in\{{\rm QNM}\}}{C_k\psi_k},
\end{equation}  
where $C_k$ represents $C^{\pm}_{\ell m n}$ and 
\begin{align} \label{eqn:definition_psi_k}
\psi_k &= e^{-i\omega_k t}\swSH{-2}{\ell{m}}{\theta,\phi}{a\omega_k}.
\end{align}
Here $\omega_k$ is one of the QNM frequency $\omega^{\pm}_{\ell mn}$. 
The set $\{\rm QNM\}$ in Eq.~\eqref{eqn:psi_fit_k} refers to a chosen collection of QNMs that we will use to fit the simulated waveform.

Since the extraction coordinates of the NR waveform discussed in this paper align with the spin of the remnant black hole, the parameter space of the remnant black hole is reduced to $\mathcal{R}=\{\delta,\chi_f\}$.  Here, the remnant mass ratio $\delta$ is defined as 
\begin{equation}\label{eqn:remnant_mass}
    \delta \equiv \frac{M_f}{M},
\end{equation}
where $M_f$ is the mass of the final remnant black hole and $M$ is the mass scale of the numerical simulation.  The magnitude of the dimensionless spin $\chi$ of the remnant BH is defined as 
\begin{equation}\label{eqn:remnant_spin}
    \chi_f \equiv \frac{J_f}{M_f^2},
\end{equation}
where $J_f$ is the magnitude of the remnant black hole's angular momentum.

\subsection{Waveform fitting} \label{subsec:Waveform_Fitting}
In this section, we will explain our multimode fitting method for the simulated signal.  The quality of the fit is described in terms of the overlap $\rho$ between the ringdown waveform $\psi_{\rm NR}$ and the fitting function $\psi_{\rm fit}$, 
\begin{align}\label{eqn:rho2}
\rho^2 = \frac{\left|\langle\psi_{\rm fit}|\psi_{\rm NR}\rangle\right|^2}
    {\langle\psi_{\rm NR}|\psi_{\rm NR}\rangle
    \langle\psi_{\rm fit}|\psi_{\rm fit}\rangle},
\end{align}
where the inner product between any two complex functions is defined as
\begin{align}\label{eqn:innerproduct}
\langle\psi_1|\psi_2\rangle \equiv
\int_{t_i}^{t_e}{dt\oint{d\Omega\psi^*_1(t,\Omega)\psi_2(t,\Omega)}}.
\end{align}
Thus, $t_i$ and $t_e$ define the initial and end times of the fit. 

The ringdown waveform $\psi_{\rm NR}$ consists of a chosen set of spherical harmonics $\{\rm NR \}$,
\begin{equation}\label{eqn:psiNR_fitting}
\psi_{\rm NR} = \frac{rh}{M}=\sum_{\{\ell{m}\}\in \{\rm NR \}}{h^{\rm{NR}}_{\ell{m}}(t)\,\swY{-2}{\ell{m}}{\theta,\phi}},
\end{equation}
where $M$ is the mass scale of the numerical simulation. Here we change the general notation $C_{\ell m}(t)$ in Eq.~\eqref{eqn:psiNR} to $h^{\rm{NR}}_{\ell{m}}(t)$ since we choose the gravitational strain $h$ to represent the gravitational wave information.

By substituting Eq.~\eqref{eqn:psi_fit_k} into Eq.~\eqref{eqn:rho2}, the overlap can be expressed as 
\begin{align}\label{eqn:rho2explicit}
\rho^2 &= \frac{\left|\sum_k{C^*_kA_k}\right|^2}{\langle\psi_{\rm NR}|\psi_{\rm NR}\rangle\sum_{i,j}{C^*_iB_{ij}C_j}},
\intertext{where}\label{eqn:Definition_A}
A_k &\equiv \langle\psi_k|\psi_{\rm NR}\rangle,
\intertext{and}\label{eqn:Definition_B}
B_{ij} &\equiv \langle\psi_i|\psi_j\rangle.
\end{align}

By extremizing $\rho^2$, we can finally express $\rho_{\rm max}^2$ in terms of $\vec{A}$ and $\mathbb B$.
\begin{equation}\label{eqn:rho2max}
\rho^2_{\rm max} = \frac1{\langle\psi_{\rm NR}|\psi_{\rm NR}\rangle}
    \vec{A}^\dag\cdot{\mathbb B}^{-1}\cdot\vec{A}.
\end{equation}
The QNM expansion coefficients in Eq.~\eqref{eqn:psi_fit_k} are then given as 
\begin{equation}\label{eqn:least-squares_coefs}
\vec{C} = {\mathbb B}^{-1}\cdot\vec{A}.
\end{equation}
A detailed derivation for extremizing $\rho^2$ is given in Ref.~\cite{Cook:2020otn}. The ringdown fitting method using Eq.~\eqref{eqn:rho2max} and Eq.~\eqref{eqn:least-squares_coefs} is referred to as the ``eigenvalue method"\cite{Cook:2020otn}.  Another quantity called mismatch ${\cal{M}}$ is used widely to show the quality of the fitting and is defined to be 
\begin{equation}\label{eqn:definition_mismatch}
{\cal{M}}=1-\rho.
\end{equation}

Explicit expressions for computing the components of $\vec{A}$ and ${\mathbb B}$ can be found in Ref.~\cite{Cook:2020otn} for the general case where extraction coordinates do not align with the spin of the remnant black hole.  The expressions for $\vec{A}$ simplify when there is alignment, and Eqs.~\eqref{eqn:Acomp+} and \eqref{eqn:Bcomp++} illustrate these simplified forms for the case of the $\omega^+$ family of QNMs.  For the components related to a quadratic QNM $(\ell_1,m_1,n_1)\times(\ell_2,m_2,n_2)$, we can simply treat the corresponding $\YSHn{\acute\ell\ell{m}n}$ as $\delta_{\acute{\ell}\ell}$, where $\ell=\ell_1+\ell_2$ and $m=m_1+m_2$.

\begin{widetext}
\begin{eqnarray}
\label{eqn:Acomp+}
\langle\psi_{\ell{m}n+}|\psi_{\rm NR}\rangle =  
    \int_{t_i}^{t_e}
    \!\!dt\,e^{i\omega^*_{\ell{m}n}t}\!\!\!\!\!\!\sum_{\{\acute\ell m\}\in\{{\rm NR}\}}
    {\!\!\!\!\!\!C_{\acute\ell m}(t)\YSHn[*]{\acute\ell\ell{m}n}
    } \\
\label{eqn:Bcomp++}
\langle\psi_{\ell{m}n+}|\psi_{\acute\ell \acute{m} \acute{n}+}\rangle = 
    \delta_{m\acute{m}}\int_{t_i}^{t_e}{
    \!\!dt\,e^{i(\omega^*_{\ell{m}n}-\omega_{\acute\ell \acute{m}\acute{n}})t}
    \sum_{\breve\ell}
    \YSHn[*]{\breve\ell\ell{m}n}
    \YSHn{\breve\ell\acute\ell \acute{m}\acute{n}}}
\end{eqnarray}
\end{widetext}

In evaluating Eqs.~\eqref{eqn:rho2max} and~\eqref{eqn:least-squares_coefs}, the inverse mode matrix ${\mathbb B}^{-1}$ is obtained as a pseudoinverse (see Sec.~2.6.2 of Ref.~\cite{numrec_c++}) by using singular value decomposition (SVD). The mode matrix ${\mathbb B}$ can be singular or nearly singular in part because there is an exponential damping term in the form of $e^{\Im[\omega_{\ell{m}n}+\omega_{\acute\ell\acute{m}\acute{n}}]t_i}$ within the integral in Eq.~\eqref{eqn:Bcomp++}. For cases where we are fitting to sets of QNMs with higher overtones, roundoff error often becomes noticeable because higher overtones decay too quickly. To decrease such roundoff error in the fitting process, we introduced a method to rescale the QNMs. 

\subsection{Rescaling the QNMs} \label{subsec: rescaling}
The process of rescaling the QNMs is straightforward.  If we wish to fit using rescaling, we simply replace each QNM $\psi_k$ with a rescaled version $\psi^\prime_k$ based on the initial time $t_i$ of the fit.
\begin{equation}\label{eqn:rescaling_psi_k_prime}
 \psi^\prime_k=e^{-\Im[\omega_k]t_i}\psi_k.
\end{equation}
With the vector $\vec{A}$ and mode matrix $\mathbb B$ rescaled in this way, a rescaled vector of expansion coefficients $\vec{C}^\prime$ will be computed by Eq.~(\ref{eqn:least-squares_coefs}) and the unscaled components can be recovered by 
\begin{equation}
C_k=e^{-\Im[\omega_k]t_i}C^\prime_k.
\end{equation}

We refer to this technique as ``rescaling", and the advantage of rescaling is its effect on the singular values of the mode matrix $\mathbb B$.  For simplicity, let $t_e$ be a time late enough in the ringdown signal that we can consider $e^{\Im[\omega_k]t_e}\to0$.  Then the rescaled matrix element
\begin{align}\label{eqn: rescaling_analytical}
\langle\psi^\prime_{\ell{m}n+}|\psi^\prime_{\acute\ell\acute{m}\acute{n}+}\rangle &= \\
\delta_{m\acute{m}}&\frac{-e^{i\Re[\omega_{\ell{m}n}-\omega_{\acute\ell\acute{m}\acute{n}}]t_i}}{i(\omega^*_{\ell{m}n}-\omega_{\acute\ell\acute{m}\acute{n}})}    \sum_{\breve\ell}
    \YSHn[*]{\breve\ell\ell{m}n}
    \YSHn{\breve\ell\acute\ell\acute{m}\acute{n}}\nonumber,
\end{align}
has removed the exponential damping term $e^{\Im[\omega_{\ell{m}n}+\omega_{\acute\ell\acute{m}\acute{n}}]t_i}$ which can dominate the magnitude of the coefficients as $t_i$ gets large.  The behavior of rescaling is explored in more depth in Appendix~\ref{app:details_rescaling}.  The example of Eq.~(\ref{eqn: rescaling_analytical}) also points out that the mode matrix coefficients can be computed without numerical integration.  However we find it is better to use the same numerical integration over time for the calculations of $\mathbb B$ as is used to compute $\vec{A}$.

\subsection{Greedy algorithm for QNM expansion coefficient} \label{subsec: greedy_algorithm_description}
In standard linear fitting, as described in Sec.~\ref{subsec:Waveform_Fitting}, the QNM expansion coefficients $C_k$ in Eq.~\eqref{eqn:psi_fit_k} are all treated as free parameters extracted simultaneously by solving Eq.~\eqref{eqn:least-squares_coefs}. Here we modify the fitting procedure with the implementation of a greedy algorithm. We sort and divide the total set of QNMs used for fitting, $\{\text{QNM}\}$, into two sets $\{\text{QNM}_f\}$ and $\{\text{QNM}_u\}$.  The expansion coefficients for the modes in $\{\text{QNM}_f\}$ are taken to have fixed, known values, while coefficients for the modes in $\{\text{QNM}_u\}$ remain as unknown quantities.  The modes are sorted so that the vector of fitting modes $\vec\psi$ with components $\psi_k$ can be written in block form as $\begin{pmatrix}^f\!\vec\psi\\^u\vec\psi\end{pmatrix}$.  Here the components of $^f\!\vec\psi$ are taken from $\{\text{QNM}_f\}$ and the components of $^u\vec\psi$ from $\{\text{QNM}_u\}$.

The vector $\vec{A}$ and mode matrix $\mathbb B$ in Eqs.~\eqref{eqn:Definition_A} and~\eqref{eqn:Definition_B} are split in a similar way to be 
\begin{align} \label{eqn:Greedy_A_Vec}
    \vec{A}&=\begin{pmatrix} 
    ^{f\!\!}\vec{A}\\
    ^{u\!}\vec{A}
    \end{pmatrix}
\intertext{and}
\label{eqn:Greedy_B_Matrix}
    \mathbb B &= \begin{pmatrix}
    ^{f}\mathbb{B} & ^c\mathbb{B}\\
    ^c\mathbb{B}^\dag & ^u\mathbb{B}
    \end{pmatrix}.
\end{align}
$^{f\!}\vec{A}$ and $^f\mathbb{B}$ are constructed from $^{f\!}\vec\psi$, while $^{u\!}\vec{A}$ and $^u\mathbb{B}$ are constructed with $^u\vec\psi$ both by using the definitions in Eqs.~\eqref{eqn:Definition_A} and \eqref{eqn:Definition_B}. 
The block matrix $^c\mathbb{B}$ has components that are inner products between the components of $^{f\!}\vec\psi$ and $^u\vec\psi$ such that
\begin{equation} \label{eqn:Definition_B_c}
    ^c\mathbb{B}_{ij}=\langle ^{f\!}\psi_i | ^u\psi_j \rangle.
\end{equation}

The overlap for the unknown QNMs $\rho_u^2$ is defined similarly to Eq.~\eqref{eqn:rho2} as
\begin{equation} \label{eqn:unknown_overlap}
\rho_u^2 = \frac{\left|\langle {}^u\psi_{\rm fit}|\psi_{\rm NR}-{}^{f\!}\psi_{\rm fit}\rangle\right|^2}
    {\langle\psi_{\rm NR}-{}^{f\!}\psi_{\rm fit}|\psi_{\rm NR}-{}^{f\!}\psi_{\rm fit}\rangle
    \langle^u\psi_{\rm fit}|^u\psi_{\rm fit}\rangle}.
\end{equation}
Here, we define the fixed, known portion of the fit function as $^f\!\psi_{\rm fit} \equiv {}^f\!\vec{C}\cdot{}^f\!\vec\psi$, and the unknown portion as $^u\!\psi_{\rm fit} \equiv {}^u\vec{C}\cdot{}^u\vec\psi$.  We will refer to $\rho_u^2$ as the partial overlap to distinguish it from the overlap in Eq.~\eqref{eqn:rho2}. The mismatch ${\cal{M}}_u$ is calculated from $\rho_u^2$ in the usual way ${\cal{M}}_u=1-\rho_u$ and is referred to as the partial mismatch. 

By extremizing the partial overlap $\rho_u^2$, the unknown QNM coefficients are found to be 
\begin{equation} \label{eqn:unknown_coefficients}
^u\vec{C} = (^u\mathbb{B})^{-1}\cdot(^{u\!}\vec{A}-^{c\!}\mathbb{B}^\dag\cdot{}^{f\!}\vec{C}).
\end{equation}

A version of our Mathematica Paclet {\tt KerrRingdown} which includes support for all of the investigations reported in this paper, including greedy fitting and quadratic modes, is freely available at~\cite{KerrRingdownCode}.

\subsection{Naming conventions}
Some conventions are defined here to help us describe further results more clearly and conveniently. During our ringdown fitting, we choose the $t_{\text{peak}}$ to be the time at which the $L^2$ norm of the gravitational strain reaches its maximum value. We then shift the simulation times so that $t_{\text{peak}}=0$.  Thus, when we specify times, such as the initial time $t_i$ and the ending time $t_e$ for the integral in Eq.~\eqref{eqn:innerproduct}, we are always referring to the time relative to $t_{\text{peak}}$. For example, $t_i=10M$ is equivalent to $t_i-t_{\text{peak}}=10M$, where $M$ is the natural mass scale for the simulation. 

For convenience, the QNMs are sometimes referred to by their frequencies, $\omega^{\pm}_{\ell mn}$. This is equivalent to naming QNMs by $(\ell,m,n,\pm)$.  Here, ``+'' designates an ``ordinary mode'', and ``-'' a ``mirror mode''. In some cases, such as in a table, sets of QNMs are listed as $\omega^{\pm}_{(\ell_1\sim\ell_2)m(n_1\sim n_2)}$ for compactness. For example, $\omega^{+}_{(2\sim3)2(0\sim1)}$ means the set of QNMs containing $\omega^{+}_{220}$, $\omega^{+}_{221}$, $\omega^{+}_{320}$, and $\omega^{+}_{321}$. The uppercase letter $N$ is used to denote the highest overtone in a set of QNMs used for fitting. For instance, we can represent a QNM set as $\omega^{+}_{(2\sim3)2(0\sim N)}$. Different signal modes $C_{\ell{m}}$ of the numerical relativity signal, Eq.~\eqref{eqn:psiNR}, are often simply referred to by their indices $(\ell,m)$. Note that we do not consider signal modes with $l>4$ in this work.

\subsection{Simulation data used for fitting} \label{subsec:simulation_data_used_for_fitting}
The work we present has been performed using the numerical simulation SXS:BBH:0305 from the Simulating eXtreme Spacetimes (SXS) catalog~\cite{SXSCatalog,Boyle:2019kee}.  Versions of this simulation have frequently been explored\cite{Giesler:2019uxc,Cook:2020otn,MaganaZertuche:2021syq,Baibhav:2023clw,Ma:2022wpv,Dhani:2020nik,Forteza:2021wfq,Mitman:2022qdl,Cheung:2022rbm}. This waveform corresponds to a binary black hole system similar to that responsible for the first gravitational wave observation, GW150914~\cite{LIGOScientific:2016aoc}. Recently, the quality of the simulated waveform has been improved by Cauchy characteristic extraction (CCE) and a mapping to the super rest frame~\cite{Moxon:2020gha,Moxon:2021gbv,Mitman:2021xkq,MaganaZertuche:2021syq,Mitman:2022kwt,Mitman:2024uss}. The version of the waveform we explore is thus in the same Bondi-Metzner-Sachs (BMS) frame where the QNMs are derived based on linear perturbation theory. 

Implementing ringdown fittings in the correct BMS frame has a number of benefits. First, the spin-weight -2 spherical harmonic modes $(2,\pm2)$, $(3,\pm2)$, and the $m=0$ modes are found to be most strongly impacted by the supertranslation to the super rest frame, and the fitting quality for these modes can be improved by using the waveform in the super rest frame~\cite{MaganaZertuche:2021syq}. Second, there is unphysical mode mixing due to the gravitational recoil if the waveform is not in the correct BMS frame~\cite{Kelly:2012nd,Boyle:2015nqa}. For example, the QNM $(2,2,0,+)$ can mix into the signal mode $(2,1)$~\cite{Ma:2022wpv} and the QNM $(3,3,0,+)$ can mix into the signal mode $(2,2)$~\cite{Cheung:2023vki}. By properly fixing the BMS frame for the waveform, we avoid these unphysical mode mixings.

In our work, we focus on fitting the signal modes $(2,2)$, $(3,2)$, and $(4,2)$, which can be impacted by the choice of the BMS frame. Therefore, we investigated a CCE waveform that corresponds to the GW150914 event and is mapped to the super rest frame. We labeled this simulation as SXS:BBH\_ExtCCE:0305 by following the naming convention in \cite{ExtCCECatalog}, although this simulation is not available on that public website yet. The remnant parameters for this simulation are $\mathcal{R}=\{0.9520177,0.6920851\}$. 

\section{Linear Fitting to SXS:BBH\_E\lowercase{xt}CCE:0305} \label{sec: Linear_Fitting}
In this section, we apply linear fitting to the waveform SXS:BBH\_ExtCCE:0305 using the ``eigenvalue method'' outlined in Sec.~\ref{sec: Method}.  To assess the robustness of the extracted QNM coefficients, we analyze distributions of the QNM expansion coefficients obtained from a series of initial fitting times, $t_i$.  We adopt a common approach in which robustness is assessed by evaluating the uncertainty of the expansion coefficients within a time window that moves across the range of initial fitting times, $t_i$. Among these windows, we select the one with the lowest uncertainty to estimate the QNM coefficients.  Details of the statistical procedures and robustness criteria are provided in Sec.~\ref{subsec:robustness_criteria}. 
Using these criteria, we conduct a thorough comparison of the QNM expansion coefficients extracted from various multimode-fitting combinations.

We demonstrate improvements in the robustness of the extracted QNM coefficients, particularly for subdominant QNMs, while fully accounting for the mode-mixing effect.  By including both signal modes and QNMs up to $\ell=4$, we find that the uncertainty in QNM $(3,2,0,+)$ decreases by a factor of $\sim2.7\times10^{-4}$ under our definition, compared to fitting only $(2\!\!\sim\!\!3,2,0\!\!\sim\!\!1,+)$ to the signal mode $(2,2)$.  We also apply the greedy algorithm outlined in Sec.~\ref{subsec: greedy_algorithm_description} to evaluate its effectiveness in enhancing the robustness of the fitted QNM coefficients.  Our findings indicate that the QNM coefficients for overtone $(2,2,2,+)$ and the subdominant mode $(3,2,1,+)$ can only be extracted robustly using the greedy algorithm. Furthermore, overtones with $n > 2$ for $(2,2,n,+)$ and $n > 1$ for $(3,2,n,+)$ cannot be extracted robustly from this dataset.  A tentative quadratic QNM, $(2,1,0,+)\times(2,1,0,+)$, in the signal mode $(4,2)$ is discussed in Sec.~\ref{subsec: nonlinear_modes_210_210}.   A summary of the robust QNMs and their contributions to the signal modes are presented in Sec.~\ref{sec:signal_residuals}.

\subsection{Criteria for assessing the robustness of the QNM expansion coefficients}\label{subsec:robustness_criteria}
In this section, we introduce detailed procedures and criteria used to assess the robustness of the QNM expansion coefficients for the linear fitting scenario.  After obtaining uncertainties for the QNM coefficients, the criteria for robustness are used to determine which QNMs can be reliably extracted from the simulated ringdown signal.  Note that some procedures are different for the nonlinear fitting scenario.

First, a chosen set of QNMs is fit to a set of signal modes.  We vary the initial time $t_i$ of the fit through a time range from $t_0\le t_i\le t_f$ to obtain the fit QNM coefficients as a function of $t_i$.  For each ringdown fit within such a multimode set, the end time $t_e$ of the fit, as defined in in Eq.~\eqref{eqn:innerproduct}, is fixed to be a time late enough that the fitting results are not sensitive to $t_e$. We set $t_e=100M$ throughout Sec.~\ref{sec: Linear_Fitting}.

In order to define a distribution of results, we specify a running time window with length of $\Delta t$ to be $[t_w,t_w+\Delta t)$, and allow $t_w$ to vary through the ranges $t_0\le t_w\le t_f-\Delta t$.  The uncertainties of the amplitudes $A^{\pm}_{\ell mn}$ and phases $\phi^{\pm}_{\ell mn}$ of the QNM expansion coefficients are calculated for the set of coefficients in each time window.  The median values of $A^{\pm}_{\ell mn}$ and $\phi^{\pm}_{\ell mn}$ in the time window with lowest uncertainty are taken as an estimate for the extracted $A^{\pm}_{\ell mn}$ and $\phi^{\pm}_{\ell mn}$.  This is similar to the choice made by Cheung \textit{et al}.\cite{Cheung:2023vki}. The details of estimating the median and calculating the uncertainty are described below. 

\begin{table*}[ht!]
\renewcommand{\arraystretch}{1.5}
\begin{tabular}{ |c|c|c| } 
 \hline
 Description&Symbol&Value  \\ [0.2cm]
 \hline 
    The length of the time window & $\Delta t$& $10M$ for overtones $n<2$ and $5M$ for overtones $n\geqslant2$\\[0.1cm]
 \hline
    The earliest initial time $t_i$ of the fitting & $t_0$ & $t_0=t_{\text{peak}}=0M$ \\[0.1cm]
\hline
    The latest initial time $t_i$ of the fitting & $t_f$ & $t_f$ depends on different cases but $t_f<t_e=100M$\\[0.1cm]
\hline
    Joint relative uncertainty of $C^{\pm}_{\ell mn}(t_i)$ for $t_w\le t_i\le t_w+\Delta t$ &$\Delta(t_w)$& Eq.~\eqref{eqn:uncertainty_QNM}\\[0.1cm]
\hline
    Lowest uncertainty among $\Delta(t_w)$ for $t_0\le t_w\le t_f-\Delta t$ &$\Delta_{min}$& Eq.~\eqref{eqn:uncertainty_QNM} at $t_{\Delta_{min}}$\\[0.1cm]
\hline
    Linear Fit Estimation for the QNM amplitude& ${}^\text{est} \! A^{\pm}_{\ell mn}$& Median $\widetilde{A}^{\pm}_{\ell mn}$ in the time window with $\Delta_{min}$\\[0.1cm]
\hline
    Linear Fit Estimation for the QNM phase& ${}^\text{est} \! \phi^{\pm}_{\ell mn}$& Median $\widetilde{\phi}^{\pm}_{\ell mn}$ in the time window with $\Delta_{min}$\\[0.1cm]
\hline
    Model Fit Median for the QNM amplitude& ${}^\text{mod} \! A^{\pm}_{\ell mn}$& Median among ${}^\text{est} \! A^{\pm}_{\ell mn}$ from fitting models\\[0.1cm]
\hline
    Model Fit Median for the QNM phase& ${}^\text{mod} \! \phi^{\pm}_{\ell mn}$& Median among ${}^\text{est} \! \phi^{\pm}_{\ell mn}$ from fitting models \\[0.1cm]
\hline

\end{tabular}
\caption{\label{tab:criteria_definition} Definitions used for evaluating the robustness of extracted $C^{\pm}_{\ell mn}$ from linear fitting.}
\end{table*}
Because data points in a single time window are limited and do not follow the normal distribution, bootstrapping~\cite{10.1214/aos/1176344552} is used to obtain more reliable estimates for the median values.   Bootstrap is a statistical technique for estimating the statistics of a dataset by sampling it with replacement. The basic procedure we follow is summarized here. 
\begin{enumerate}
    \item The original dataset contains the $N_0$ values of $A^{\pm}_{\ell mn}$ or $\phi^{\pm}_{\ell mn}$ for each $t_i$ in the time window $[t_w,t_w+\Delta t)$. 
    \item We choose the number of bootstrap samples to be 100\,000, with each sample containing $N_0$ data points. 
    \item We draw the data points for each sample randomly, with replacement, from the original dataset. Then we find the median of each sample. 
    \item The bootstrapped median of each sample's median is computed and we denote this median appropriately as either $\widetilde{A}^{\pm}_{\ell mn}$ or $\widetilde{\phi}^{\pm}_{\ell mn}$. 
    \item The uncertainty of the bootstrapped median is taken as the range of the $95\%$ two-sided confidence interval for the samples' medians. We denote the uncertainty of the amplitude as $\delta_{\!A}$, and the uncertainty of the phase as $\delta_{\!\phi}$.
\end{enumerate}
After bootstrapping, the joint relative uncertainty of a specific QNM expansion coefficient in the time window $[t_w,t_w+\Delta t)$ is defined as
\begin{equation}\label{eqn:uncertainty_QNM}
    \Delta(t_w)=\sqrt{\left(\frac{\delta_{\!A}}{\widetilde{A}}\right)^2+\left(\frac{\delta_{\!\phi}}{\widetilde{\phi}}\right)^2}.
\end{equation}
Here we omit the subscription $\ell mn$ for $A$ and $\phi$ for the sake of compactness. 

This joint relative uncertainty $\Delta(t_w)$ is calculated for each time window $t_0\le t_w\le t_f-\Delta t$. The time window with the lowest uncertainty $\Delta_{min}$ is taken as the time range in which the QNM coefficients are fit most robustly.  Within this time window, the bootstrapping medians $\widetilde{A}^{\pm}_{\ell mn}$ and $\widetilde{\phi}^{\pm}_{\ell mn}$ are the estimates for the QNM expansion coefficients for this fitting model. The estimated QNM amplitude and phase are denoted as ${}^\text{est} \! A^{\pm}_{\ell mn}$ and ${}^\text{est} \! \phi^{\pm}_{\ell mn}$.  We will refer to these estimates as Linear Fit Estimates, and denote the start of the time window with the lowest uncertainty as $t_{\Delta_{min}}$.

Our basic robustness criterion for each estimated QNM expansion coefficient is the following.  A QNM coefficient is considered robust if: 
\begin{RC}
    The lowest uncertainty $\Delta_{min}$ is smaller than 0.01.  The length of the time window $\Delta t$ is set to $10M$ for the QNMs with $n\leqslant1$ and is set to $5M$ for the QNMs with $n\geqslant2$.  We do this because higher overtones damp quickly and only contribute to the signal for a short time.
\end{RC}
The value for this empirical criterion, $\Delta_{min} < 0.01$, was chosen based on observed instabilities in the plots of $A^{\pm}_{\ell mn}$ and $\phi^{\pm}_{\ell mn}$ versus $t_i$ for cases with $\Delta_{min} > 0.01$, but is clearly subjective.

In practice, different fitting models can produce estimated QNM coefficients that satisfy the criterion above. To obtain a robust final estimate, we take the median of the linear fit estimates over these different fitting models.  We refer to these values as the Model Fit Median.  Each such estimated QNM amplitude and phase is denoted as ${}^\text{mod} \! A^{\pm}_{\ell mn}$ and ${}^\text{mod} \! \phi^{\pm}_{\ell mn}$.  The various symbols and definitions defined above are summarized in Table~\ref{tab:criteria_definition}.

To estimate the uncertainty in the Model Fit Median, we use the following approach. For each fitting model contributing to the Model Fit Median, we compute the $95\%$ confidence intervals for the distributions of $A^{\pm}_{\ell mn}$ and $\phi^{\pm}_{\ell mn}$ within their respective lowest-uncertainty windows. The overall uncertainty is then determined by identifying the smallest lower bound and largest upper bound among these confidence intervals across all fitting models. These bounds provide a conservative estimate of the uncertainty, capturing the full range of potential variations across the different models.

\subsection{Signal residuals} \label{sec:signal_residuals}
Using greedy linear fitting, we have determined the set of QNMs which can be robustly fit to the $h^{\rm{NR}}_{22}$, $h^{\rm{NR}}_{32}$, and $h^{\rm{NR}}_{42}$ modes of the simulated signal designated SXS:BBH\_ExtCCE:0305 based on the criteria in Sec.~\ref{subsec:robustness_criteria}.  The model fit medians for the robust modes are listed in Table~\ref{tab: C220_Summary_Table}.  Note that no retrograde modes were robustly fit using our procedures, and $C^+_{321}$ does not fully meet our robustness criteria.  Details about how we obtained these values are given in the following sections.  Before we consider these details, it is useful to consider how well these robust modes represent the original signal.
\begin{table}[ht!]
\renewcommand{\arraystretch}{1.5}
\begin{tabular}{ |c|c|c| } 
 \hline
 QNM  &${}^\text{mod} \! A^{\pm}_{\ell mn}$& ${}^\text{mod} \! \phi^{\pm}_{\ell mn}$    \\ [0.2cm]
\hline 
    $C^{+}_{220}$& $0.97100^{+0.00005}_{-0.00005}$&$1.482222^{+0.000090}_{-0.000024}$\\[0.2cm]
\hline
    $C^{+}_{320}$& $0.037814^{+0.000014}_{-0.000024}$&$-0.8288^{+0.0004}_{-0.0011}$\\[0.2cm]
\hline
    $C^{+}_{420}$& $0.00200^{+0.00005}_{-0.00007}$&$-2.307^{+0.035}_{-0.032}$\\[0.2cm]
\hline 
    $C^{+}_{221}$& $4.224^{+0.031}_{-0.070}$&$-0.658^{+0.007}_{-0.030}$\\[0.2cm]
\hline 
    $C^{+}_{222}$& $11.92^{+0.17}_{-0.50}$&$2.762^{+0.04}_{-0.024}$\\[0.2cm]
\hline 
    $C^{+}_{321}$& 0.268&-2.930\\[0.2cm]
\hline 
\end{tabular}
\caption{Summary of robust QNM coefficients.  These values of ${}^\text{mod} \! A^{\pm}_{\ell mn}$ and ${}^\text{mod} \! \phi^{\pm}_{\ell mn}$ were utilized to generate the residual plots presented in Sec.~\ref{sec:signal_residuals}.} \label{tab: C220_Summary_Table}
\end{table}

\begin{figure}[htb!]
\centering
\includegraphics[width=\columnwidth]{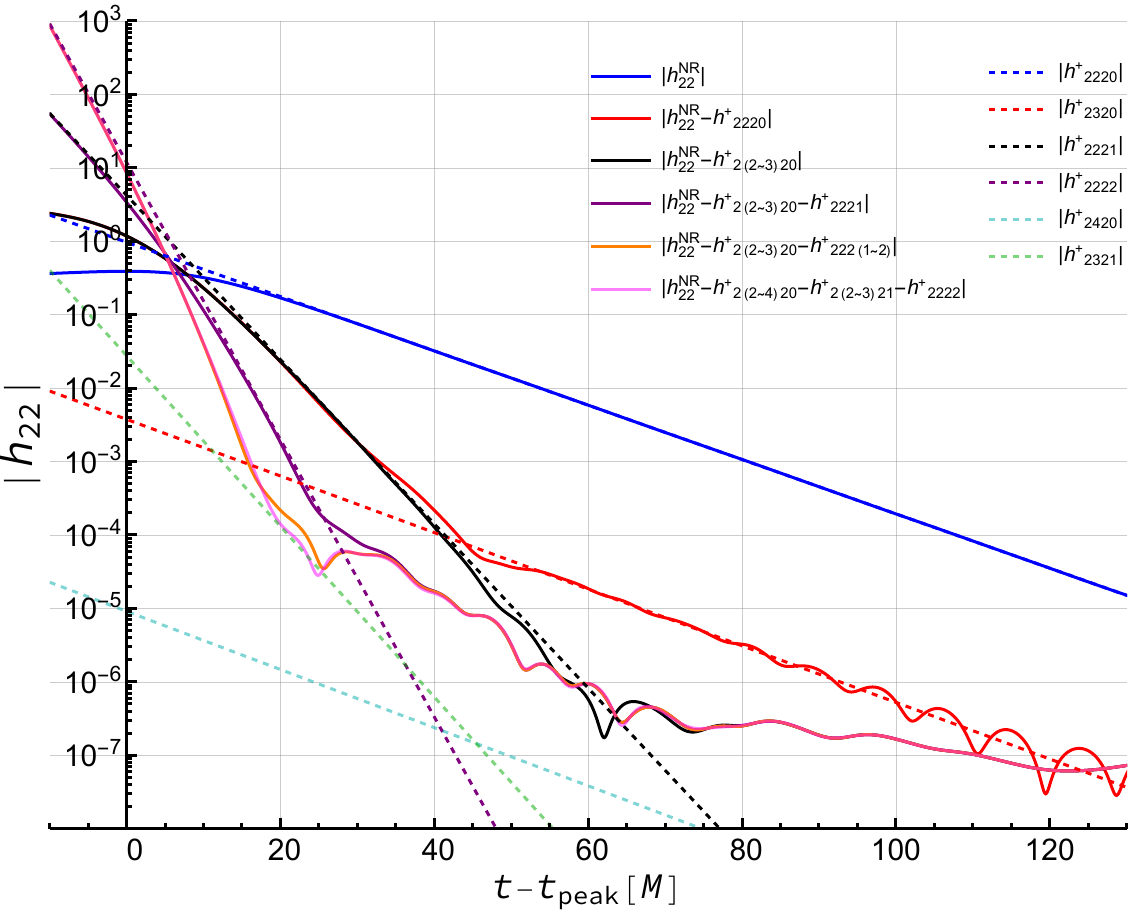}
\caption{The magnitudes of various parts of the $h^{\rm{NR}}_{22}$ are plotted as solid lines after combinations of $h^+_{2\ell2n}$ are subtracted from it. The magnitudes of various $h^+_{2\ell2n}$ are plotted as dashed lines. The color of each $|h^+_{2\ell2n}|$ is set to the color of the most closely matching residual of $h^{\rm{NR}}_{22}$, in which $|h^+_{2\ell2n}|$ is its dominant contribution.}
\label{fig:h22SubtractQNM}
\end{figure}

\begin{figure}[htb!]
\centering
\includegraphics[width=\columnwidth]{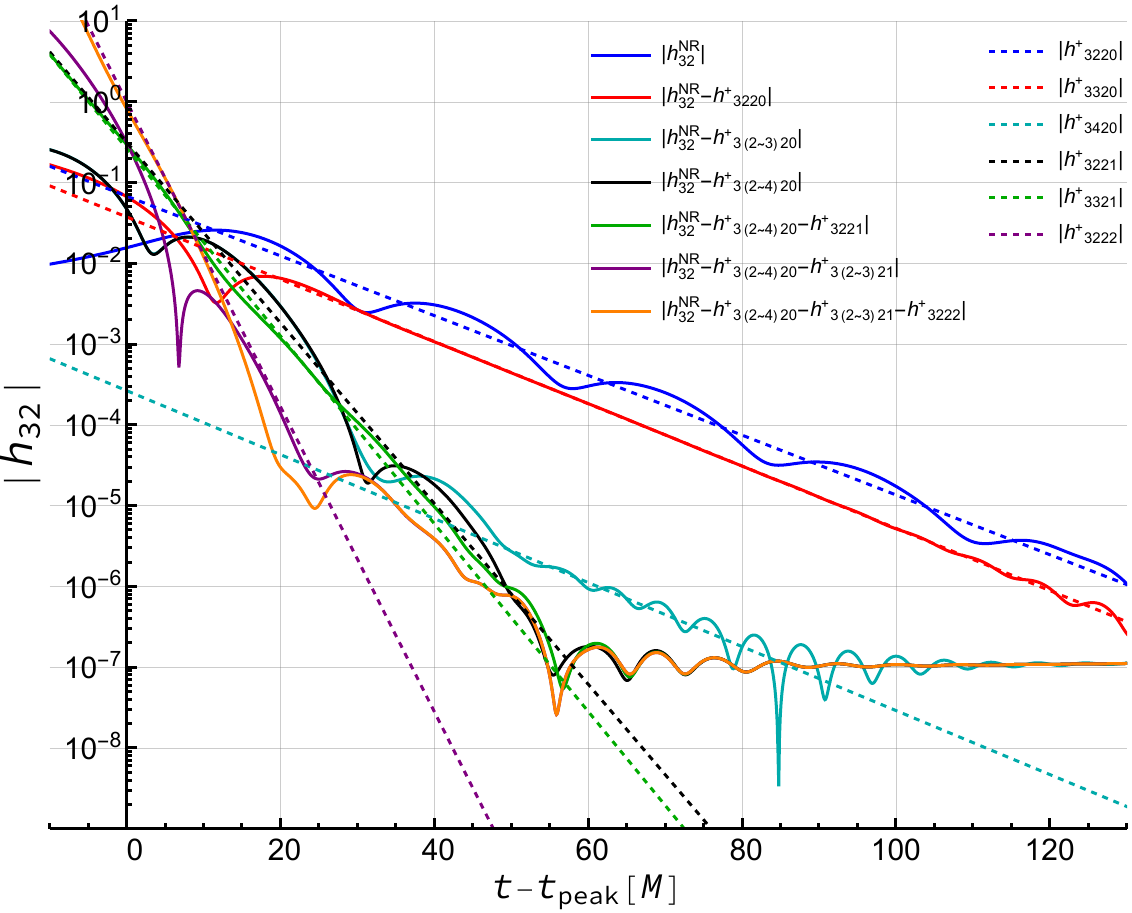}
\caption{The magnitudes of various parts of the $h^{\rm{NR}}_{32}$ are plotted as solid lines after combinations of $h^+_{3\ell2n}$ are subtracted from it. The magnitudes of various $h^+_{3\ell2n}$ are plotted as dashed lines. The color of each $|h^+_{3\ell2n}|$ is set to the color of the most closely matching residual of $h^{\rm{NR}}_{32}$, in which $|h^+_{3\ell2n}|$ is its dominant contribution.}
\label{fig:h32SubtractQNM}
\end{figure}

\begin{figure}[htb!]
\centering
\includegraphics[width=\columnwidth]{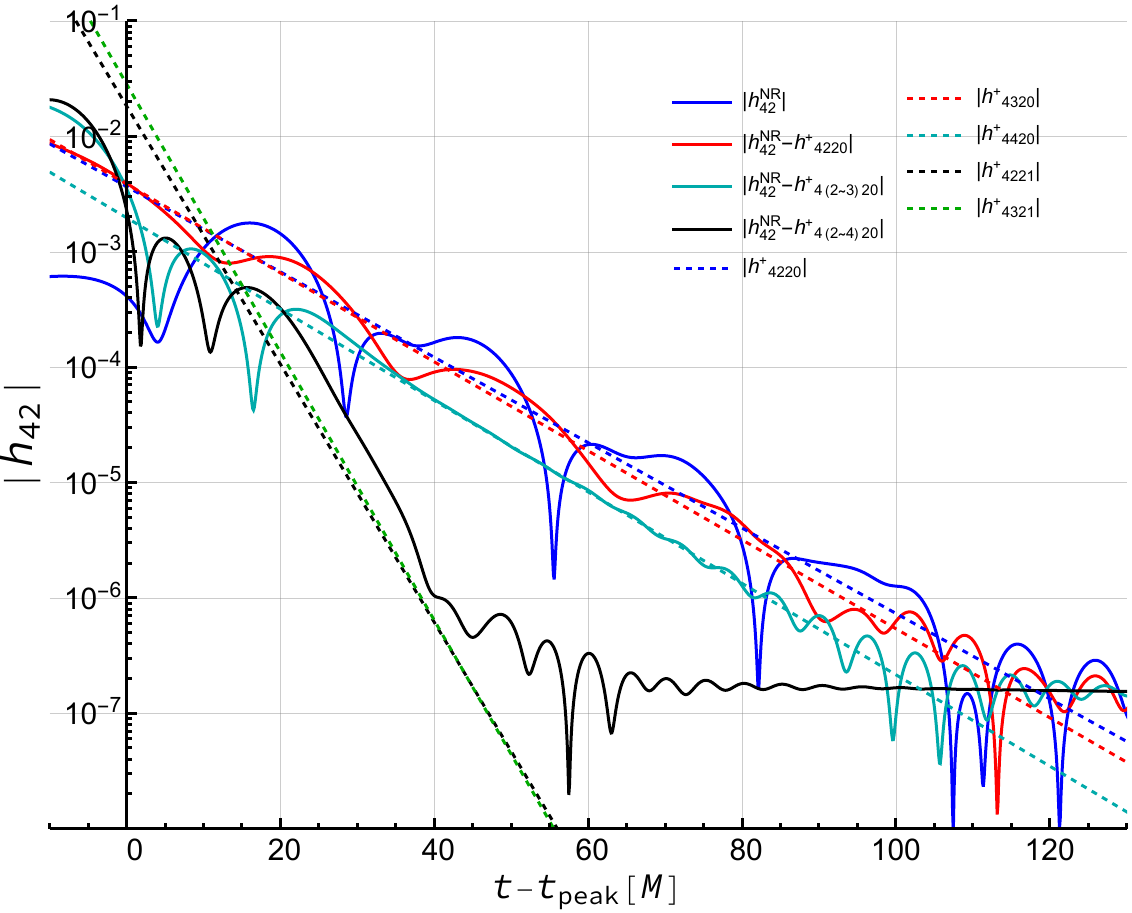}
\caption{The magnitudes of various parts of the $h^{\rm{NR}}_{42}$ are plotted as solid lines after combinations of $h^+_{4\ell2n}$ are subtracted from it. The magnitudes of various $h^+_{4\ell2n}$ are plotted as dashed lines.  The color of each $|h^+_{4\ell2n}|$ is set to the color of the most closely matching residual of $h^{\rm{NR}}_{42}$, in which $|h^+_{4\ell2n}|$ is its dominant contribution.}
\label{fig:h42SubtractQNM}
\end{figure}

Each waveform expansion coefficient $h_{\ell m}$ in Eq.~\eqref{eqn:psiNR_fitting} can be expanded in terms of the QNM expansion coefficients $C^{\pm}_{\ell mn}$ as
\begin{equation} \label{eqn: hlm_expand_Clmn}
\begin{split}
h_{\acute{\ell}m} = & \sum_{\ell n} {\{ C^{+}_{\ell mn}e^{-i \omega_{\ell mn}t} \mathcal{A}_{\acute{\ell}\ell mn}} \\
 & +(-1)^{\ell+\acute{\ell}}C^{-}_{\ell mn}e^{i \omega^*_{\ell(-m)n}t} \mathcal{A}^*_{\acute{\ell}\ell(-m)n}\}.
\end{split}
\end{equation}
For the prograde modes which we consider here, we define the contribution of a specific QNM $(\ell, m, n, +)$ to specific $h_{\acute{\ell} m}$ to be $h^+_{\acute\ell\ell mn}$, 
\begin{equation} \label{eqn: QNM_contribution_signal}
    h^+_{\acute\ell\ell mn}(t) = C^{+}_{\ell mn}e^{-i \omega^+_{\ell mn}(t-t_{\text{peak}})} \mathcal{A}_{\acute{\ell}\ell mn}. 
\end{equation}
Here we use the QNM coefficients in Table~\ref{tab: C220_Summary_Table} for the $C^{+}_{\ell mn}$.  In Figs.~\ref{fig:h22SubtractQNM}, \ref{fig:h32SubtractQNM}, and \ref{fig:h42SubtractQNM}, we present residual plots where we display the magnitudes of the full signal, the full signal with the contributions of various QMNs subtracted (yielding fit residuals), and the individual $h^+_{\acute\ell\ell mn}(t)$.

In Fig.~\ref{fig:h22SubtractQNM}, we present the residual plot for $h^{\rm{NR}}_{22}$.  The solid colored lines represent the full signal and the residuals of $h^{\rm{NR}}_{22}$ after successively subtracting various $h^+_{2\ell 2n}$ components.  The magnitudes of different $h^+_{2\ell 2n}$ contributing to $h^{\rm{NR}}_{22}$ are plotted as dashed colored lines.  For ease of comparison, we match the color of each $|h^+_{2\ell 2n}|$ to that of the corresponding residual of $h^{\rm{NR}}_{22}$ in which $|h^+_{2\ell 2n}|$ is its dominant contribution.  In Fig.~\ref{fig:h32SubtractQNM} and~\ref{fig:h42SubtractQNM}, we present similar plots for signals $h^{\rm{NR}}_{32}$ and $h^{\rm{NR}}_{42}$.

In Figs.~\ref{fig:h22SubtractQNM}, \ref{fig:h32SubtractQNM}, and \ref{fig:h42SubtractQNM}, a superposition of $h^+_{(2\sim4)20}$ is clearly seen to be the primary contribution to each of the signal modes $h^{\rm{NR}}_{(2\sim4)2}$ in the late ringdown stage, for times $t \geq 50M$.  These results highlight the necessity of including signal modes and QNMs up to $\ell=4$ to account for mode-mixing effects.   After subtracting the $h^+_{\acute\ell(2\sim4)20}$ contributions from the signal, the numerical noise settles around $10^{-7}$ at late times for $h^{\rm{NR}}_{(3\sim4)2}$, indicating that the QNM coefficients $C^+_{(2\sim4)20}$ are extracted with high accuracy.

The higher overtones in the signal decay rapidly.  For instance, the $h^+_{(2\sim3)222}$ contributions drop below a magnitude of $10^{-4}$ before $t = 30M$. This confirms that, as expected, overtones with $n > 1$ are typically significant only in the early stages, when the ringdown signal likely includes significant nonlinear effects.  Conventional wisdom is to assume that linear perturbation theory may not accurately model the ringdown near the signal peak, potentially leading to poor fits for these higher overtones.\footnote{Reference~\cite{Giesler:2024hcr} provides new evidence that, with sufficiently accurate waveforms and by including quadratic modes, perturbation theory may accurately model the ringdown near the signal peak.}  The residual signal $|h^{\rm{NR}}_{22} - h^+_{2(2\sim3)20} - h^+_{222(1\sim2)}|$ in Fig.~\ref{fig:h22SubtractQNM} shows significant unmodeled effects at times beyond $t = 20M$. We attribute this to a combination of inaccuracies in the subtracted $h^+_{2\ell2n}$, the presence of additional QNMs, quadratic modes, and other nonlinear effects.

In Fig.~\ref{fig:h42SubtractQNM}, we do not plot the residual signals $|h^{\rm{NR}}_{42} - h^+_{4(2\sim4)20} - h^+_{4221}|$ and $|h^{\rm{NR}}_{42} - h^+_{4(2\sim4)20} - h^+_{4(2\sim3)21}|$, as the contribution of $h^+_{4(2\sim3)21}$ to $h^{\rm{NR}}_{42}$ overlaps with a potential nonlinear mode $(2,1,0,+) \times (2,1,0,+)$.  The details are discussed in Sec.~\ref{subsec: nonlinear_modes_210_210} and displayed in Fig.~\ref{fig:h42SubtractQNMNL42}.

\subsection{Selection of fitting models and implementing the greedy algorithm}
In Sec.~\ref{sec: Method}, we gave a detailed description of our fitting methods for a given fitting model, where we define a fitting model as a specific combination of a set of signal modes and a set of fitting QNMs.  Here we address two additional aspects of our fitting procedures: the selection of fitting models, and the order of iterative QNM coefficient extraction using the greedy algorithm.

\subsubsection{The importance of multimode fitting}\label{subsubsec: the_importance_of_multimode_fitting}
In constructing fitting models, we consistently account for the full mode-mixing effect by including signal modes and QNMs from $2 \leq\ell\leq 4$.  In this work, we focus entirely on the $m=2$ axial modes.  Specifically, our models fit QNMs $(2\!\sim\!4, 2, 0\!\sim\!N)$ to the signal modes $(2\!\sim\!4, 2)$.  Under this framework, the primary difference among models lies in the highest overtone number within the QNM set.  Below, we present a representative example to illustrate the importance of considering mode-mixing effects.

\begin{figure}[htb!]
\centering
\includegraphics[width=\columnwidth]{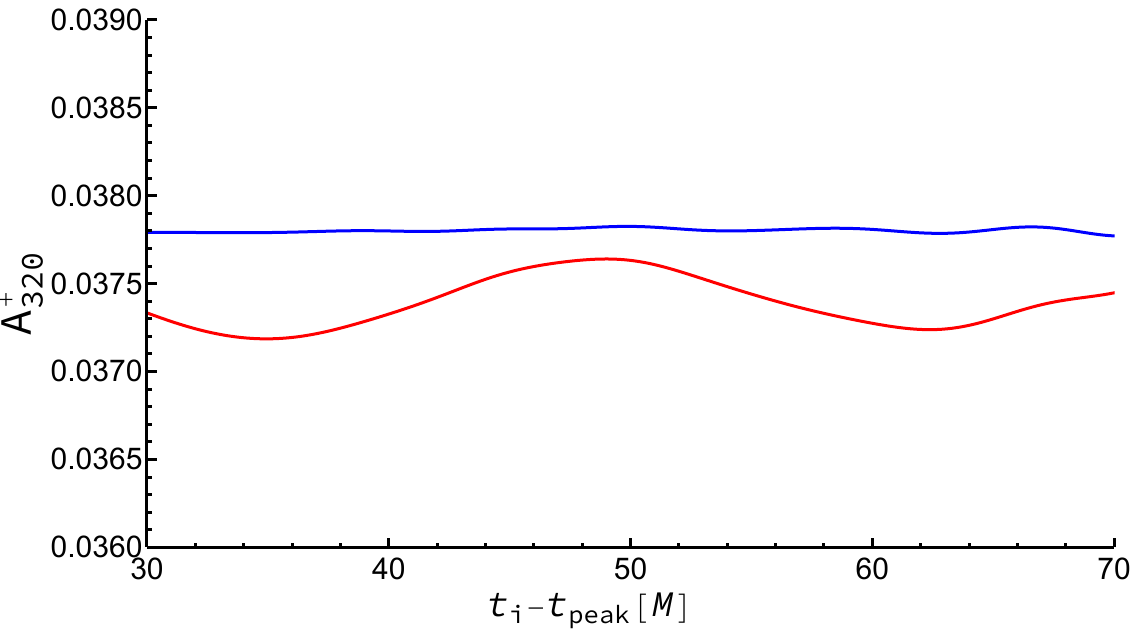} 
\caption{The amplitude $A^+_{320}$ is plotted as a function of the initial fitting time $t_i$. The blue line presents model No. 1 in Table~\ref{tab:C320_Linear_Fitting}, while the red line presents model No. 4.  The difference is that model No. 4 omits the $(4,2)$ signal mode and the QNMs $(4,2,0\!\sim\!1,+)$ from the fit model.}
\label{fig:A320CompareModeMixing}
\end{figure}

Table~\ref{tab:C320_Linear_Fitting} and Fig.~\ref{fig:A320CompareModeMixing} illustrate the importance of simultaneously fitting multiple signal modes and multiple QNMs.  We presents the linear fit estimations for the QNM coefficient $C^+_{320}$ across several fitting models.  In Table~\ref{tab:C320_Linear_Fitting}, models No. 1 and No. 2 include $\ell=2\sim4$ signal modes, models No. 3 and No. 4 omit the $\ell=4$ signal mode, and model No. 5 uses only the $\ell=2$ signal mode.  
Comparing models No. 1 with Nos. $3\!\sim\!4$, we observe that $\Delta_{min}$ decreases by approximately two order of magnitude when both signal mode $(4,2)$ and QNMs $(4,2,n,+)$ are included in the multimode fitting. 
This enhancement in the robustness of the QNM coefficient $C^+_{320}$ is further illustrated in Fig.~\ref{fig:A320CompareModeMixing}, which plots $A^+_{320}$ as a function of $t_i$ for models No. 1 and No. 4 from Table~\ref{tab:C320_Linear_Fitting}.  The amplitude $A^+_{320}$ from model No. 4 (red line) exhibits oscillatory behavior across $t_i$, whereas the amplitude from model No. 1 (blue line) shows much less variation.  Even though the contribution of $h^+_{3420}$ to $h^{\rm{NR}}_{32}$ is roughly $2$ orders of magnitude smaller than the contributions from $h^+_{3220}$ and $h^+_{3320}$ (see Fig.~\ref{fig:h32SubtractQNM}), omitting the contributions of the $(4,2)$ signal mode to mode-mixing effects leads to considerable oscillation in the extracted value of $A^+_{320}$ for model No. 4.  For model No. 5, where only signal mode $(2,2)$ is included, the coefficient $C^+_{320}$ exhibits a significant loss of accuracy, with $\Delta_{min}$ failing to meet the basic robustness criterion~1.

In Table~\ref{tab:C320_Linear_Fitting}, also note that between models No. 1 and No. 2 the difference is the omission of QNMs $(4,2,0\!\sim\!1,+)$ in model No. 2.  There is a similar difference between models No. 3 and No. 4.  While omitting the $\ell=4$ QNM does make a difference in the results, it is the inclusion of the $(4,2)$ signal mode which has the most significant effect on the robustness of $C^+_{320}$.

\begin{table}[ht!]
\begin{tabular}{ |c|c|c|c|c|c|c| } 
 \hline
 No. &Signals&QNMs &${}^\text{est} \! A^+_{320}$ &${}^\text{est} \! \phi^+_{320}$ & $t_{\Delta_{min}}/M$ & $\Delta_{min}$   \\ [0.2cm]
\hline
    1& $(2\!\sim\!4,2)$& {$\omega^+_{(2\sim4)2(0\sim1)}$}&0.03780&-0.8286&36.9&$4.0\times10^{-5}$ \\[0.2cm]
\hline
   2&$(2\!\sim\!4,2)$ &{$\omega^+_{(2\sim3)2(0\sim1)}$}&0.03775&-0.8277&55.7&$3.5\times10^{-4}$ \\[0.2cm]
\hline
   3&$(2\!\sim\!3,2)$ &{$\omega^+_{(2\sim4)2(0\sim1)}$}&0.03763&-0.8278&16.1&$1.1\times10^{-3}$ \\[0.2cm]
\hline
    4&$(2\!\sim\!3,2)$&{$\omega^+_{(2\sim3)2(0\sim1)}$}&0.03723&-0.8296&29.1&$1.0\times10^{-3}$ \\[0.2cm]
\hline
   5&$(2,2)$ &{$\omega^+_{(2\sim3)2(0\sim1)}$}&0.07073&-3.1297&58.2&$0.15$ \\[0.2cm]
\hline
\end{tabular}
\caption{\label{tab:C320_Linear_Fitting} QNM amplitude $A^+_{320}$ and phase $\phi^+_{320}$ extracted from different fitting models. Each row represents a different fitting model, which is a specific combination of a set of signal modes (column 2) and a set of fitting QNMs (column 3).  The definitions for ${}^\text{est} \! A^+_{320}$, ${}^\text{est} \! \phi^+_{320}$, $t_{\Delta_{min}}$, and $\Delta_{min}$ are listed in Table~\ref{tab:criteria_definition}. }
\end{table}

The improvement in the robustness of QNM coefficients with multimode fitting is evident across most QNMs listed in Table~\ref{tab: C220_Summary_Table}.  While certain overtones, such as $(2,2,2,+)$, show no significant improvement from the inclusion of the signal mode $(4,2)$ and the QNMs $(4,2,n)$, their robustness does not degrade with this choice of the fitting model.

\subsubsection{Extracting QNM coefficients with the greedy algorithm} \label{subsubsec: order_to_extract_greedy}
When using the greedy algorithm to iteratively extract QNM coefficients, we extract the fundamental modes ($n=0$) first, and then proceed to higher overtones, incrementing one overtone order at a time. This ordering is chosen because lower overtones, which have longer decay times as shown in Figs.~\ref{fig:h22SubtractQNM}, \ref{fig:h32SubtractQNM}, and \ref{fig:h42SubtractQNM}, can be extracted more robustly and accurately than higher overtones.

A secondary question is whether QNM coefficients with the same overtone value should be extracted simultaneously or successively, in order from the dominant to subdominant QNMs. Using the greedy algorithm, we find that fixing the dominant QNM coefficients does not generally improve the robustness of the subdominant QNM coefficients with the same overtone value. To illustrate, we use the subdominant QNM $(3,2,1,+)$ as a representative example.

In Table~\ref{tab:C321_Compare_Model_Boot_Median}, we compare the robustness of the extracted $C^+_{321}$ coefficient by applying the greedy algorithm in three different ways.  In this example, the fitting model includes signal modes $(2\sim4,2)$ and the prograde QNMs $\omega^+_{(2\sim4)2(0\sim N)}$, where $N=1\sim2$.  In the first case, all QNMs are fit simultaneously.  In the second case, all the $n=0$ QNMs are fixed to the values given in Table~\ref{tab: C220_Summary_Table}, and all the other QNMs are fit simultaneously.  For the third case, in addition to fixing the $n=0$ QNMs, we also fix $C^+_{221}$ to the value given in Table~\ref{tab: C220_Summary_Table}, and the remaining QNMs are fit simultaneously.  For the fitting models with $N=1$, we clearly see that fixing only the $n=0$ QNMs improves the robustness of the subdominant QNM $(3,2,1,+)$, while fixing one of the $n=1$ QNMs negates the benefit of the greedy algorithm. The fitting models with $N=2$ display different patterns and will be discussed further in Sec.~\ref{subsec: compare_linear_models}.  While the robustness of the subdominant QNM $(3,2,1,+)$ does not benefit from the greedy algorithm, fixing one of the $n=1$ QNMs still gives the worst uncertainty $\Delta_{min}$.

Interestingly, we do not see the same effect in the fundamental modes.  Fixing $C^+_{220}$ and then fitting for the remaining $n=0$ QNMs has a neutral effect on the robustness of the subdominant QNMs $(3\sim4,2,0,+)$.

\begin{table}[ht!]
\renewcommand{\arraystretch}{1.5}
\begin{tabular}{ |c|c|c|c|c|c| } 
\hline
    No.&Signals&QNMs &${}^\text{est} \! A^+_{321}$ &${}^\text{est} \! \phi^+_{321}$ & $\Delta_{min} $ \\ [0.1cm]
\hline
        1&$(2\!\sim\!4,2)$&$\omega^+_{(2\sim4)2(0\sim 1)}$ &$0.484$&$-3.039$ & $0.018$ \\[0.1cm] 
\hline
        $1^*$&$(2\!\sim\!4,2)$&$\omega^+_{(2\sim4)2(0\sim 2)}$ &$0.431$&$-2.996$ & $0.015$ \\[0.1cm] 
\hline
        2&$(2\!\sim\!4,2)$& \begin{tabular}{@{}c@{}}$\omega^+_{(2\sim4)2(0\sim 1)}$ \\ ${}^{f} C^{+}_{(2\sim4)20}$\end{tabular}&$0.268$&$-2.930$ & $0.007$  \\[0.1cm] 
\hline
        $2^*$&$(2\!\sim\!4,2)$& \begin{tabular}{@{}c@{}}$\omega^+_{(2\sim4)2(0\sim 2)}$ \\ ${}^{f} C^{+}_{(2\sim4)20}$\end{tabular}&$0.452$&$-3.033$ & $0.019$  \\[0.1cm] 
\hline
        3&$(2\!\sim\!4,2)$& \begin{tabular}{@{}c@{}}$\omega^+_{(2\sim4)2(0\sim 1)}$ \\ ${}^{f} C^{+}_{(2\sim4)20}, {}^{f}C^+_{221}$\end{tabular}&$0.457$&$-3.004$ & $0.024$\\[0.1cm]
\hline
        $3^*$&$(2\!\sim\!4,2)$& \begin{tabular}{@{}c@{}}$\omega^+_{(2\sim4)2(0\sim 2)}$ \\ ${}^{f} C^{+}_{(2\sim4)20}, {}^{f}C^+_{221}$\end{tabular}&$0.276$&$-2.854$ & $0.029$\\[0.1cm]
\hline
\end{tabular}
\caption{\label{tab:C321_Compare_Model_Boot_Median} Comparison of three different applications of the greedy algorithm to the same fitting model used to extract $C^+_{321}$.  In the first case, the greedy algorithm is not used.  In the second, $C^{+}_{220}$, $C^{+}_{320}$, and $C^{+}_{420}$ are fixed while the other QNMs are simultaneously fit.   In the third, $C^+_{221}$ is also held fixed.  For all three cases, we present results from using $N=1$ and $N=2$ to set the highest included overtone. The asterisk in the model number denotes fitting models with $N=2$, to distinguish them from those with $N=1$.}
\end{table}
 
Empirically, we find that it is usually beneficial, and never detrimental, to fix all robustly fit QNMs with the same overtone value, when using the greedy algorithm.  The mathematical basis for this is that the spin-weighted spheroidal harmonics for QNMs with the same overtone value form a minimal set\cite{London:2020uva} and encode distinct mode information.  In contrast, modes with the same $\ell$ and $m$ but different overtone values do not.  We summarize the implementation of the greedy algorithm as follows: when extracting QNMs with the same overtone value $n$ simultaneously, all robust QNMs with overtone values less than $n$ are fixed. The process begins with $n=0$ and proceeds iteratively to higher overtones, increasing $n$ by one at each step.

\subsection{Robustness of QNM coefficients across fitting models} \label{subsec: compare_linear_models}
The robustness criterion described in Sec.~\ref{subsec:robustness_criteria} is based on a commonly used approach, where the robustness of the QNM coefficients is assessed by evaluating the uncertainty within a specific extraction window.  This criterion works effectively for a single fitting model. In this section, we address an additional consideration:  when a QNM coefficient can be extracted robustly by multiple fitting models, should final determination of a coefficient's robustness also depend on the consistency of the extracted value across the different fitting models?

There are multiple ways to construct different fitting models.  However, we will consider the specific case where only the number of included overtones is allowed to vary.  Giesler \textit{et~al}.\cite{Giesler:2019uxc} were the first to show that including higher overtones could improve the quality of a ringdown fit, and push the initial fitting time $t_i$ which could yield a good fit to very early in the ringdown signal.  Our conjecture is that, at least within some range of overtones, a set of robustly fit QNMs should yield consistent values if the fits are truly robust.  We explore this conjecture using a representative example where $C^+_{221}$ is extracted using linear fitting both without and with application of the greedy algorithm.
\begin{figure}[htb!]
\centering
\includegraphics[width=\columnwidth]{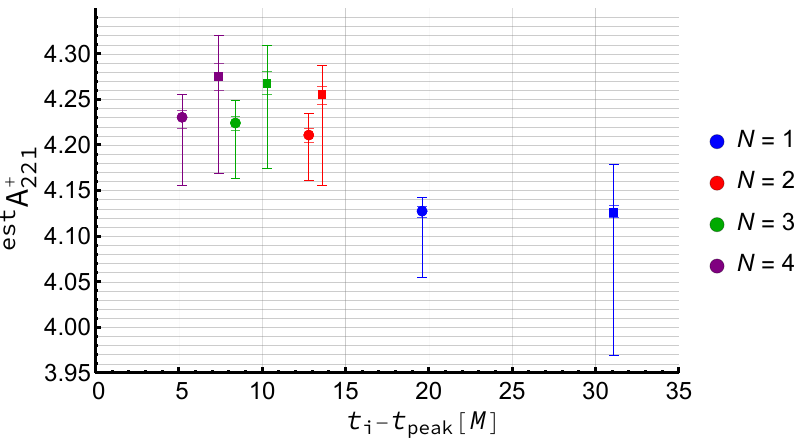}
\caption{The linear fit estimates for ${}^\text{est} \! A^+_{221}$ obtained by fitting QNMs $\omega^+_{(2\sim4)2(0\sim N)}$ to signal modes $(2\sim4,2)$. Each value is plotted at the time $t_{\Delta_{min}}$ marking the start of the extraction window.  Squares indicate fits performed without applying the greedy algorithm, while circles denote fits in which $C^{+}_{(2\sim4)20}$ are fixed to their values in Table~\ref{tab: C220_Summary_Table}.  Each point includes two error bars.  The larger bar represents the $95\%$ confidence interval of $A^+_{221}$ over $t_i \in [t_{\Delta_{min}}, t_{\Delta_{min}} + 10M)$, while the smaller bar denotes the median uncertainty (this would be $\delta_A$ in Eq.~\eqref{eqn:uncertainty_QNM}).}
\label{fig:A221BestExtractedL4}
\end{figure}

In Fig.~\ref{fig:A221BestExtractedL4}, each dot represents a pair $\{t_{\Delta_{min}},{}^\text{est} \!A^+_{221}\}$, estimated by fitting QNMs $\omega^+_{(2\sim4)2(0\sim N)}$ to the signal modes $(2\!\!\sim\!\!4,2)$.  To vary the fitting model, we only change the maximum overtone $N$ to include in the model.  Fit values obtained with greedy fitting are displayed as circles, while the simple linear-fit values are displayed as squares.  Each set of points includes two error bars.  The larger errors indicate the $95\%$ confidence interval for the distribution of values for $A^+_{221}$ within the lowest uncertainty window, while the smaller error range is the standard deviation of the bootstrapped medians.  In this example, all fitting models satisfy the basic robustness criterion, though the corresponding ${}^\text{est} \!A^+_{221}$ across models exhibit variations larger than their respective error bars.  The ${}^\text{est} \!A^+_{221}$ for $N=2\sim4$ demonstrate good consistency, while the $N=1$ models shows noticeable deviation. This behavior is not unexpected.  When only QNMs $\omega^+_{(2\sim4)2(0\sim 1)}$ are included in the fitting, unmodeled contributions from higher overtones, especially $n=2$, influence the extraction of $C^+_{221}$. Although some of these higher overtones cannot be fit robustly, their inclusion effectively absorbs residual contributions and mitigates other noise in the signal.  Therefore, when calculating model fit medians, we omit models where the highest overtone $N$ matches the overtone value $n$ of the targeted QNM coefficient. For the examples shown in Fig.~\ref{fig:A221BestExtractedL4}, models with $N=1$ are excluded from the model fit median calculation.

\begin{figure}[htb!]
\centering
\includegraphics[width=\columnwidth]{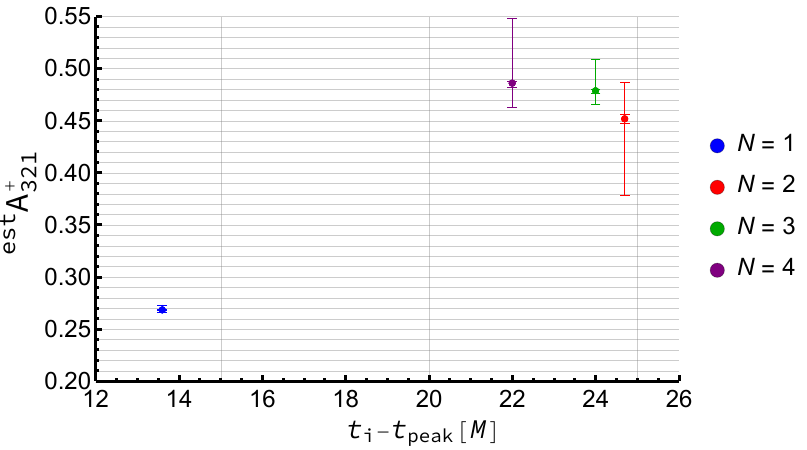}
\caption{The linear fit estimates for ${}^\text{est} \! A^+_{321}$ obtained by fitting QNMs $\omega^+_{(2\sim4)2(0\sim N)}$ to signal modes $(2\sim4,2)$ with $C^{+}_{(2\sim4)20}$ held fixed.  See Fig.~\ref{fig:A221BestExtractedL4} for additional details.}
\label{fig:A321BestExtractedL4}
\end{figure}

Figure~\ref{fig:A321BestExtractedL4} shows the interesting behavior for ${}^\text{est} \! A^+_{321}$.  For this case, only the values obtained from fitting models with $N=1$ meet the basic robustness criterion.  Additionally, ${}^\text{est} \! A^+_{321}$ shows poor consistency across fitting models with $N=1$ to $4$, and we find that the extraction window for $N=1$ is before the times for the $N>1$ cases.  Typically, we expect to find that the extraction window shifts to earlier times as $N$ increases.  Although we record the values of $C^+_{321}$ in Table~\ref{tab: C220_Summary_Table}, we consider it only a tentatively robust QNM.  Furthermore, $C^+_{321}$ is not held fixed in subsequent iterations of the greedy algorithm.

In conclusion, we consider the consistency of extracted QNMs within an extraction window, basic robustness criterion~1 from Sec.~\ref{subsec:robustness_criteria}, to be a necessary but not sufficient criteria for an extracted QNM to be considered robust.  We propose the additional criteria that:
\begin{RC}
    An extracted QNM is only considered robust if it satisfies criterion~1 and corresponds to a model for which the largest overtone $N$ is larger than the overtone $n$ of the extracted QNM.  Furthermore, if there are multiple such extracted values, then we use the model fit median values ${}^{\text{mod}}A^{\pm}_{\ell{m}n}$ and ${}^{\text{mod}}\phi^{\pm}_{\ell{m}n}$ as the robust extracted values for use in the greedy algorithm.  Finally, we take the uncertainty of a model fit median as the total range of the $95\%$ confidence intervals for all included models.
\end{RC}

\subsection{Robustness and the greedy algorithm}\label{subsec:robustness_and_the_greedy_algorithm}
The greedy algorithm described in Sec.~\ref{subsec: greedy_algorithm_description} has two key effects on the linear fitting process. First, fixing the coefficients of lower overtones during fitting reduces the uncertainties in the extracted QNM coefficients. Second, it enhances consistency across fitting models. These two improvements are more evident for higher overtones, such as the (2,2,2,+) overtone.

\begin{figure}[htb!]
\centering
\includegraphics[width=\columnwidth]{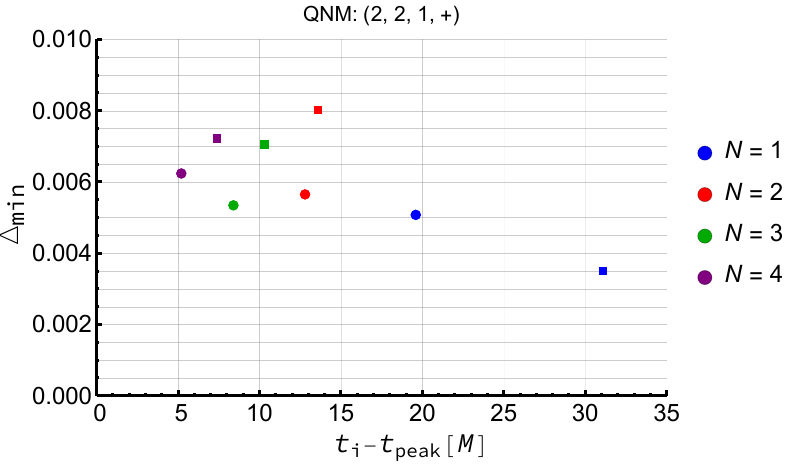}
\caption{Values of $\Delta_{min}$ for linear fit estimates of ${}^\text{est} \! A^+_{221}$ are obtained by fitting QNMs $\omega^+_{(2\sim4)2(0\sim N)}$ to signal modes $(2\sim4,2)$.  See Fig.~\ref{fig:A221BestExtractedL4} for additional details.}
\label{fig:Delta221BestExtractedL4}
\end{figure}

In Fig.~\ref{fig:A221BestExtractedL4}, the error bars for the extracted $A^+_{221}$ are noticeably reduced following the implementation of the greedy algorithm. Figure~\ref{fig:Delta221BestExtractedL4} offers a direct comparison of $\Delta_{min}$ across each fitting model. When the $C^{+}_{(2\sim4)20}$ are fixed during the fitting, $\Delta_{min}$ decreases for models with $N=2\sim4$.

\begin{figure}[htb!]
\centering
\includegraphics[width=\columnwidth]{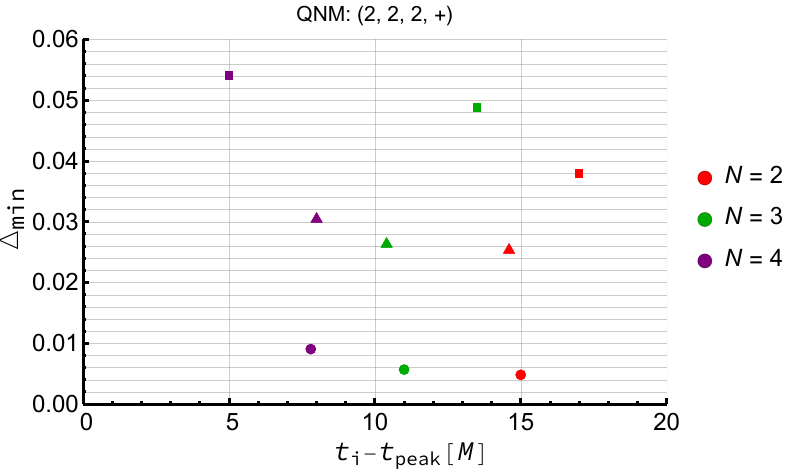}
\caption{Values of $\Delta_{min}$ for linear fit estimates of ${}^\text{est} \! A^+_{222}$ obtained by fitting QNMs $\omega^+_{(2\sim4)2(0\sim N)}$ to signal modes $(2\sim4,2)$.  Each value is plotted at the time $t_{\Delta_{min}}$ marking the start of the extraction window.  Squares indicate fits performed without applying the greedy algorithm.  Circles and triangles denote greedy fits with two different sets of fixed QNMs.  In both cases, $C^{+}_{(2\sim4)20}$ are held fixed, and circles denote the results when $C^+_{221}$ is also held fixed.}
\label{fig:Delta222BestExtractedL4}
\end{figure}

\begin{figure}[htb!]
\centering
\includegraphics[width=\columnwidth]{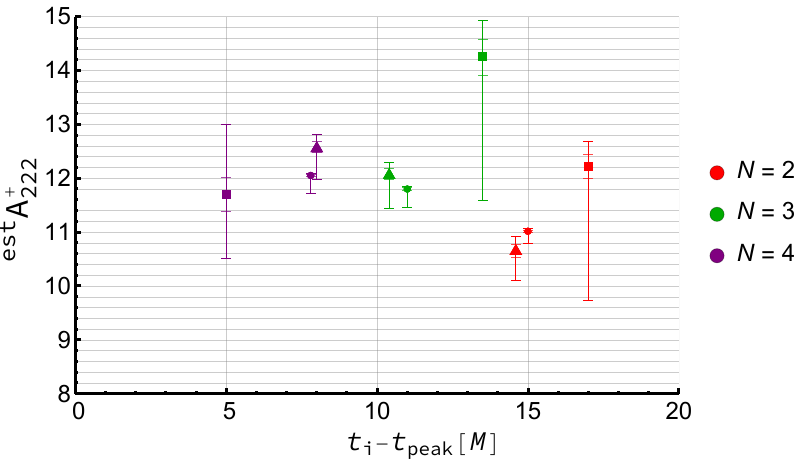}
\caption{Linear fit estimates for ${}^\text{est} \! A^+_{222}$ obtained by fitting QNMs $\omega^+_{(2\sim4)2(0\sim N)}$ to signal modes $(2\sim4,2)$.  See Figs.~\ref{fig:A221BestExtractedL4} and \ref{fig:Delta221BestExtractedL4} for additional details.}
\label{fig:A222BestExtractedL4}
\end{figure}

The improvement in the robustness of the extracted $C^+_{222}$ is more pronounced than that of $C^+_{221}$, as shown in Fig.~\ref{fig:Delta222BestExtractedL4} and~\ref{fig:A222BestExtractedL4}. In Fig.~\ref{fig:Delta222BestExtractedL4}, we observe that $\Delta_{min}$ of $C^+_{222}$ decreases progressively with each successive iteration of the greedy algorithm. When comparing $\Delta_{min}$ values from fits with both $C^{+}_{(2\sim4)20}$ and $C^+_{221}$ fixed to those without the greedy algorithm, we find that $\Delta_{min}$ is reduced by approximately an order of magnitude and the extracted $C^+_{222}$ values for $N=2\sim4$ all satisfy our robustness criteria.  In Fig.~\ref{fig:A222BestExtractedL4}, the variation in the extracted $C^+_{222}$ values across different $N$ also diminish with each iteration. We conclude that the QNM coefficient (2,2,2,+) can only be considered robust when the greedy algorithm is applied.

\subsection{Potential nonlinear mode}\label{subsec: nonlinear_modes_210_210}

\begin{figure}[htb!]
\centering
\includegraphics[width=\columnwidth]{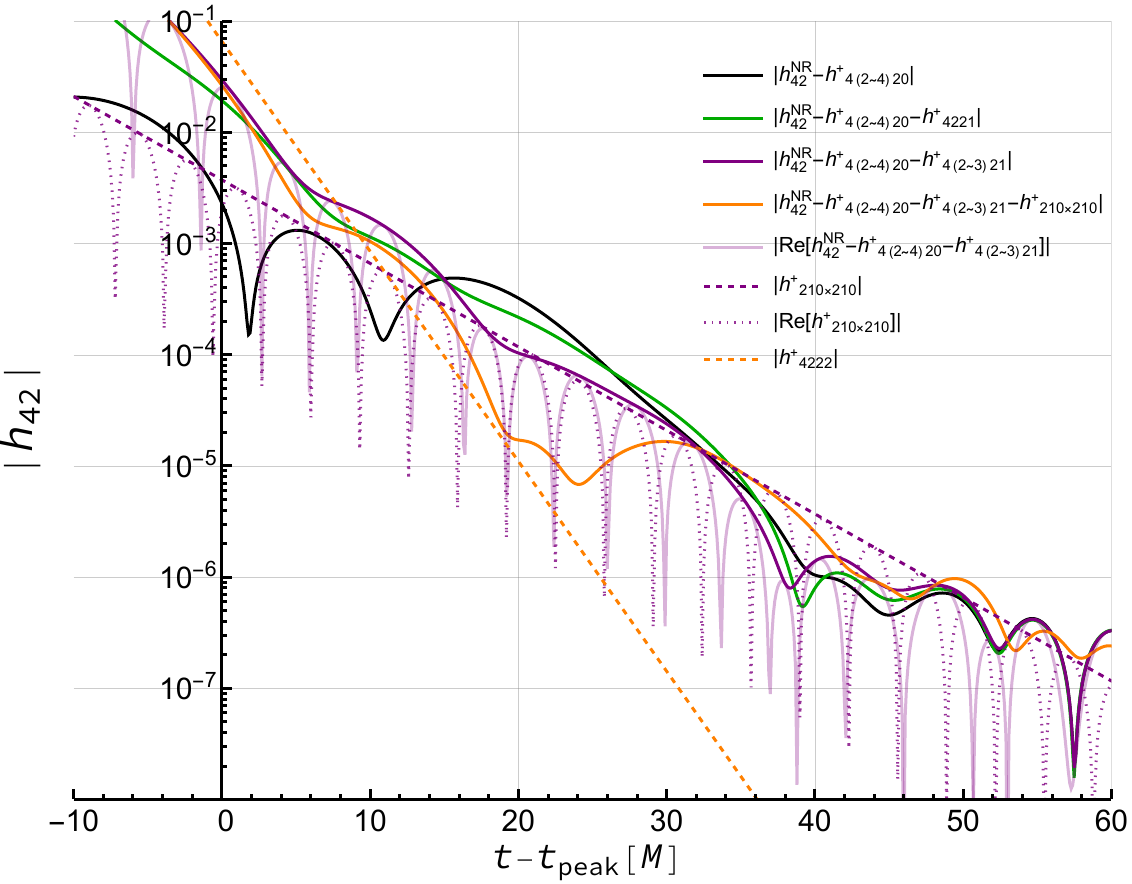}
\caption{The magnitudes of various parts of the $h^{\rm{NR}}_{42}$ are plotted as solid lines after combinations of $h^+_{4\ell2n}$ are subtracted from it. The magnitude of the potential quadratic QNM $h^+_{210\times210}$ is plotted as a purple dashed line.  The faint purple line is plotted as the magnitudes of the residual $\Re[h^{\rm{NR}}_{42} - h^+_{4(2\sim4)20} - h^+_{4(2\sim3)21}]$, and the dotted purple line is the magnitude of $\Re[h^+_{210\times210}]$.}
\label{fig:h42SubtractQNMNL42}
\end{figure}

In Fig.~\ref{fig:h42SubtractQNMNL42}, we include the residual $|h^{\rm{NR}}_{42} - h^+_{4(2\sim4)20}|$ shown in Fig.~\ref{fig:h42SubtractQNM} and now include the previously omitted residuals, $|h^{\rm{NR}}_{42} - h^+_{4(2\sim4)20} - h^+_{4221}|$ and $|h^{\rm{NR}}_{42} - h^+_{4(2\sim4)20} - h^+_{4(2\sim3)21}|$.

After we subtract only $h^+_{4(2\sim4)20}$ from $h_{42}$ in Fig.~\ref{fig:h42SubtractQNMNL42}, there is a consistent slope from around $t=20M$ to $t=35M$ in the residual $|h^{\text{NR}}_{42}-h^+_{4(2\sim4)20}|$. We find that its slope is not a single QNM but a combination of several different modes.  One part is the superposition of $h^+_{4221}$ and $h^+_{4321}$. After we subtract both, the residual $|h^{\text{NR}}_{42}-h^+_{4(2\sim4)20}-h^+_{4(2\sim3)21}|$ in Fig.~\ref{fig:h42SubtractQNMNL42} shows another consistent slope. 

We find that a tentative quadratic QNM\cite{Mitman:2022qdl,Cheung:2022rbm} can contribute to this remnant signal. The quadratic QNM $(2,1,0,+)\times(2,1,0,+)$ has a similar damping rate to the purple line over $t=18M\sim30M$.  By comparing the $|\Re[h^{\text{NR}}_{42}-h^+_{4(2\sim4)20}-h^+_{4(2\sim3)21}]|$ to the $|\Re[h^+_{4(210\times210)}]|$, we see that their frequencies are also consistent around $t=18M\sim25M$.  In Fig.~\ref{fig:h42SubtractQNMNL42}, the purple dashed and dotted lines are for the contribution of the quadratic QNM $|h^+_{4(210\times210)}|$ and $|\Re[h^+_{4(210\times210)}]|$ where $C^+_{210\times210}$ is obtained from a linear-greedy fit. The extracted amplitude and phase of the quadratic QNM $(2,1,0,+)\times(2,1,0,+)$ are $\{0.0037, 1.02\}$.  The fitting model we used here includes the QNMs $\omega^+_{22(0\sim 2)}$, $\omega^+_{32(0\sim 1)}$, $\omega^+_{420}$, and the quadratic QNM $\omega^+_{210\times210}$ fit to signal modes $(2\!\sim\!4,2)$\footnote{Note that the quadratic QNM $\omega^+_{210\times210}$ only contributes to signal mode $(4,2)$ during fitting, while mode-mixing for other QNMs is fully taken into account.}.  The QNMs $C^+_{(2\sim4)20}$ and $C^+_{(2\sim3)21}$ were fixed in the greedy method by using values from Table~\ref{tab: C220_Summary_Table}.  A careful choice of the QNM set included in the fitting is necessary for obtaining a reasonable estimate for $C^+_{210\times210}$.  Note that the $\Delta_{min}=0.034$ for the quadratic QNM in this fitting model, which means that it fails our robustness criterion~1.  However, we did not find any other QNMs that could possibly contribute to this part of the signal.  Since the ringdown waveform might already have significant nonlinear effects at $t=20M$, this might explain why it is so difficult to extract the coefficient of the subdominant QNM $(4,2,1,+)$. 

\section{Robustness and Nonlinear Fitting}\label{sec:nonlinear_fitting}
In Sec.~\ref{sec: Linear_Fitting}, the robustness of QNM coefficients extracted via linear fitting was explored both with and without incorporating the greedy method.  One key aspect of evaluating the robustness of each QNM made use of a set of QNM values acquired within an extraction window.  This set of QNMs was the foundation of robustness criterion 1 defined in Sec.~\ref{subsec:robustness_criteria}.  In essence, the most consistent set of QNM values was used to define the best extraction window, and the same measure of consistency was then used to decide if the QNM values within that window were considered robust.  It would be prudent to consider an alternative approach for obtaining a set of QNM values.

With linear fitting, the properties of the remnant black hole are treated as known quantities.  On the other hand, nonlinear ringdown fitting treats them as unknowns which are determined as part of the fit\cite{Cook:2020otn}.  Each nonlinear fit, parametrized by its initial time $t_i$, is treated as a function of the remnant parameters $\delta$ and $\chi_f$ [see Eqs.~(\ref{eqn:remnant_mass}) and (\ref{eqn:remnant_spin})].  Then, the value of the mismatch ${\mathcal M}$ is minimized over the two-dimensional parameter space $\mathcal{P}=\{\delta,\chi_f\}$,\footnote{The remnant parameter space is constrained such that $0<\delta<1$ and $0<\chi_f<1$.} yielding the estimated values $\delta_{\text{est}}$ and ${\chi_f}_\text{est}$ along with the linear fit values for the QNM coefficients for the estimated values of the remnant parameters.

In this section, we consider using consistency between the remnant parameters estimated by nonlinear fitting, and those obtained directly from the numerical simulation, as an alternative approach for obtaining a distribution of QNM values.  We will take  
\begin{equation} \label{eqn:nonlinear_parameter_error}
    \epsilon\equiv \sqrt{\left(\frac{\delta M_f}{M}\right)^2+(\delta \chi_f)^2},
\end{equation}
as our measure of the error in the estimated parameters.  Here, $\delta M_f/M$ is the difference between the $\delta(=M_f/M)$ from the simulation and our estimated $\delta_{\text{est}}$, and $\delta \chi_f$ is the difference between the $\chi_f$ from the simulation and our estimated ${\chi_f}_\text{est}$.

Our primary objective in using nonlinear fitting is not to evaluate the accuracy of remnant properties across different fitting cases, but to use nonlinear fitting, with the greedy method, as an alternative way to assess the uncertainty in extracted QNM coefficients and identify robust QNMs.  We start with regular nonlinear fitting by extracting the remnant parameters and QNM expansion coefficients as a function of the initial fit time $t_i$.  From this, we can obtain the error in the estimated remnant parameters as a function of $t_i$ for each fitting mode we consider.  Our central assumption is that the QNM coefficients are most accurately obtained when they are extracted where $\epsilon$ is small.  There will be a range of initial times $t_i$ over which $\epsilon$ is small, and this can be used to define an initial set of QNM values.  A key point is that each element of this set will have its own values for the remnant parameters $\delta_{\text{est}}$ and ${\chi_f}_\text{est}$ in addition to values for the $C^\pm_{\ell{m}n}$.

The greedy method naturally extends to nonlinear fitting, but there is one subtlety which must be considered.  For a fixed QNM with coefficient ${}^fC^\pm_{\ell{m}n}$, when its contribution to the signal is evaluated, the values of the remnant parameters are important since the arguments of the exponentials such as in Eqs.~(\ref{eqn:Acomp+}) and (\ref{eqn:Bcomp++}) take the form 
\begin{align}
    i\omega_{\ell{m}n}t &=i\left[M_f\omega_{\ell{m}n}(\chi_f)\right](t/M)/\delta,
\end{align}
where $t/M$ is the dimensionless simulation time.  For linear fitting, all QNMs are evaluated with the same values of the remnant parameters.  However our implementation of greedy fitting allows for each fixed QNM to be defined with its own, independent values for the remnant parameters.  This flexibility allows for many possible approaches to implementing nonlinear-greedy fitting, with some approaches being considerably more expensive computationally than others.

In short, we will use the greedy method in two distinct ways.  Our greedy approach is an iterative method where, during the $n$th iteration, we are attempting to extract the robust $n$th overtones.  At the end of the $n$th iteration, we will have accumulated for each fitting model, a set of fit values where each element of the set contains fit values for the remnant parameters and all of the QNM expansion coefficients with overtone $n$.  From each set, we can determine which QNMs can be considered robust for each fitting mode.  To do this, we construct a joint relative uncertainty $\Delta_{NL}$ for each QNM with overtone $n$ for each set.  The joint relative uncertainty $\Delta_{NL}$ is the same as $\Delta(t_w)$ from Eq.~(\ref{eqn:uncertainty_QNM}) except that the elements of the set have been selected based on different criteria.  We will consider a particular QNM to be robust based on the following:
\begin{RC}
    An extracted QNM is considered robust when there is at least one set with uncertainty $\Delta_{NL}<0.01$ from a model for which the largest overtone $N$ is larger than the overtone $n$ of the extracted QNM.
\end{RC}

Unlike our approach for linear fitting, we do not construct model-fit median values for the amplitude and phase for the robust modes.  Instead, we combine the elements of the individual sets for each model which satisfies robustness criterion~3, into a single robust set.  This robust set is then used to construct final fit values for the amplitudes and phases of the robust QNMs and the extracted remnant parameters.

During the $i$th iteration of the greedy method, based on the robust sets of QNMs, we will create sets of nonlinear fitting results for each fitting model which will be held fixed during the current nonlinear greedy solve.  To reduce the computational cost, we limit this to only using distributions of robust QNMs from the $(i-1)$th iteration.  This means that we construct a correlated distribution for the robust modes from the robust set from iteration $i-1$, and randomly select values for the ${}^fC^\pm_{\ell{m}(i-1)}$.  However, the values for the ${}^fC^\pm_{\ell{m}n}$ with $n<i-1$ are fixed to median values from their robust sets.  For simplicity, the values of $\delta$ and $\chi_f$ associated with {\em all} fixed ${}^fC^\pm_{\ell{m}n}$ are set to their median values from the robust set from iteration 0.  

We will discuss additional details of our nonlinear-greedy approach below.  For those not interested in the details, we find that the QNM coefficients estimated through nonlinear-greedy fitting are consistent with those from linear-greedy fitting but exhibit larger uncertainties.

\subsection{Initialization} \label{subsec: regular_nonlinear_fitting}
\begin{figure}[htb!]
\centering
\includegraphics[width=\columnwidth]{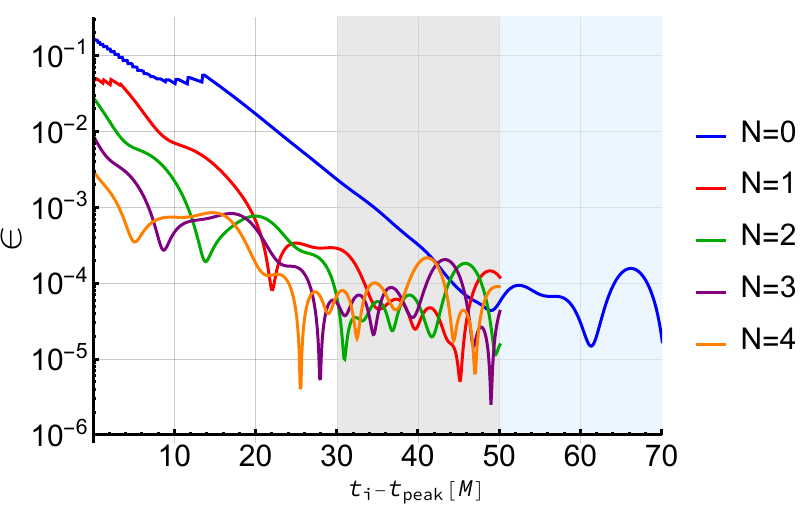}
\caption{Errors in the estimated remnant parameters are shown as a function of $t_i$ for models including QNMs up to the highest overtone $N=0\!\sim\!4$. The models fit QNMs $\omega^+_{(2\sim4)2(0\sim N)}$ to signal modes $(2\!\sim\!4,2)$. The light blue band ($t_i  \in [50M,70M]$) indicates the plateau of low remnant errors for $N=0$, while the light gray band ($t_i \in [30M,50M]$) corresponds to $N=1\!\sim\!4$.}
\label{fig:errorIte0NLFL4}
\end{figure}

Nonlinear fitting begins with all QNM coefficients in a given fitting model taken as unfixed.  For consistency with our exploration of linear fitting, we used nonlinear fitting models that fit the QNMs $\omega^+_{(2\sim4)2(0\sim N)}$ to signal modes $(2\!\sim\!4,2)$, where $N=0\!\sim\!4$.  In Fig.~\ref{fig:errorIte0NLFL4}, we plot the errors $\epsilon(t_i)$ for the estimated remnant parameters obtained from nonlinear fitting for each model.
For each model, we wish to extract a set of fit values.  To do this, we choose a window across $t_i$ within which the $\epsilon(t_i)$ has attained a roughly consistent small value.  

During this initial iteration, we will determine values for the fundamental modes $C^+_{(2\sim4)20}$.  As with linear fitting, we will only include data from models with $N>0$ when determining final median values.  The gray shaded region of Fig.~\ref{fig:errorIte0NLFL4} shows the common window $30M\le t_i\le50M$ used for the $4$ models with $N>0$.  For the model with $N=0$, we extended the range for $t_i$ out to $70M$ and used a different window with $50M\le t_i\le70M$.

From each window, we extracted 201 fit results which constitute the initial set of results for each model.  Using bootstrapping, as described in Sec.~\ref{subsec:robustness_criteria} with $N_0=201$, we construct bootstrap medians (analagous to the linear fit estimation) for the amplitudes and phases of $C^+_{(2\sim4)20}$.  The results for each model are presented in Table~\ref{tab:C220nonlinear_Bootstrap_Median_L4}.
\begin{table}[htb!]
\begin{tabular}{ |c|c|c|c|c|c|c| } 
 \hline
 Models &$A^+_{220}$ &$\phi^+_{220}$ & $A^+_{320}$ &$\phi^+_{320}$& $A^+_{420}$ &$\phi^+_{420}$   \\ [0.1cm]
\hline
     $N=0$&$0.97138$&$1.48187$&$0.03782$&$-0.8289$&$0.001995$&$-2.303$ \\ [0.1cm]
\hline
    $N=1$&$0.97104$&$1.48305$&$0.03780$&$-0.8282$&$0.001978$&$-2.289$ \\ [0.1cm]
\hline
    $N=2$&$0.97124$&$1.48303$&$0.03781$&$-0.8276$&$0.001986$&$-2.305$  \\ [0.1cm]
\hline
    $N=3$&$0.97129$&$1.48296$&$0.03781$&$-0.8274$&$0.001985$&$-2.281$ \\ [0.1cm]
\hline
    $N=4$&$0.97139$&$1.48267$&$0.03781$&$-0.8277$&$0.001984$&$-2.306$ \\ [0.1cm]
\hline
\end{tabular}
\caption{\label{tab:C220nonlinear_Bootstrap_Median_L4} Estimated QNM amplitudes and phases for $C^+_{(2\sim4)20}$ are shown for models fitting $C^+_{(2\sim4)2(0\sim N)}$ to signal modes $(2\!\sim\!4,2)$. The extracted QNM coefficients are estimated as bootstrapped medians over a $20M$ time window: $t_i \in [50M,70M]$ for $N=0$ and $t_i \in [30M,50M]$ for $N=1\!\sim\!4$. }
\end{table}
For ease of comparison, the estimated amplitudes and phases $\{A^+_{\ell mn}, \phi^+_{\ell mn}\}$ from linear fitting from Table~\ref{tab:C220nonlinear_Bootstrap_Median_L4} are: $C^+_{220}=\{0.97100,1.48222\}$, $C^+_{320}=\{0.03781,-0.8288\}$, and $C^+_{420}=\{0.00200,-2.307\}$.  We find good consistency between the nonlinear models and the linear estimates.

\begin{figure*}[htbp]
    \centering
    \begin{subfigure}[t]{0.32\textwidth}
        \centering
        \includegraphics[width=\columnwidth]{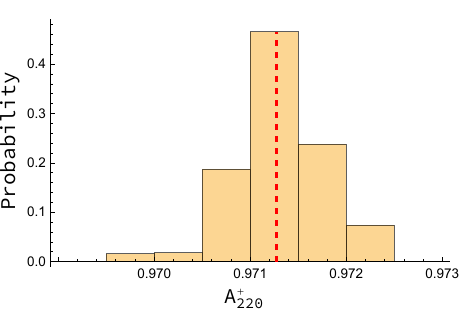}
        \label{fig:A220}
    \end{subfigure}
    \hfill
    \begin{subfigure}[t]{0.32\textwidth}
        \centering
        \includegraphics[width=\columnwidth]{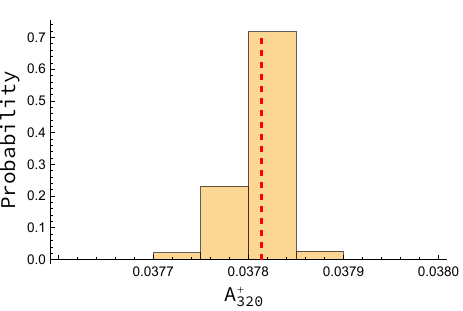}
        \label{fig:A320}
    \end{subfigure}
    \hfill
    \begin{subfigure}[t]{0.32\textwidth}
        \centering
        \includegraphics[width=\columnwidth]{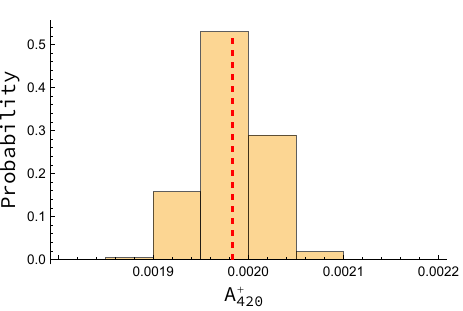}
        \label{fig:A420}
    \end{subfigure}
    \caption{Probability distributions of $A^{+}_{(2\sim4)20}$ derived from 500 correlated configurations of $C^+_{(2\sim4)20}$, generated from the multivariate Gaussian copula distribution obtained in the zeroth iteration.}
    \label{fig:Hist_Amp_Fundamental_Random}
\end{figure*}

Combining the sets from models with $N=1\sim4$, we created an $804$ element robust set.  To create a smaller and more uniformly sampled set of QNMs for subsequent iterations of the greedy method, we generated $500$ element correlated sets for each $C^+_{(2\sim4)20}$ by sampling from a multivariate Gaussian copula distribution, based on the robust set, using the \texttt{CopulaDistribution} function in Mathematica.  Table~\ref{tab: C220_Random_value} present the median values for each QNM, along with their $95\%$ confidence intervals.  Figure~\ref{fig:Hist_Amp_Fundamental_Random} presents the probability histograms of the amplitudes $A^+_{(2\sim4)20}$ from this distribution.
\begin{table}[ht!]
\renewcommand{\arraystretch}{1.5}
\begin{tabular}{ |c|c|c| } 
 \hline
 QNM  &$A^{+}_{\ell 20}$& $\phi^{+}_{\ell 20}$    \\ 
\hline 
    $C^{+}_{220}$&$0.9713^{+0.0009}_{-0.0010}$&$1.4828^{+0.0011}_{-0.0012}$\\
\hline
    $C^{+}_{320}$&$0.03781^{+0.00004}_{-0.00006}$&$-0.8278^{+0.0016}_{-0.0032}$\\
\hline
    $C^{+}_{420}$&$0.00198^{+0.00006}_{-0.00007}$&$-2.29^{+0.04}_{-0.04}$\\
\hline 
\end{tabular}
\caption{Median values and $95\%$ confidence intervals for the amplitudes and phases of $C^+_{(2\sim4)20}$ obtained from $500$ samples of the correlated distribution created during the initialization step.} \label{tab: C220_Random_value}
\end{table}

While we will use the resampled correlated sets of the $C^+_{(2\sim4)20}$ as the basis for subsequent greedy iterations, we use the original $804$-element robust sets to generate bootstrapped median values for $C^+_{(2\sim4)20}$ and the remnant parameters.  Table~\ref{tab: C220_bootstrap_value} presents these values.  The $95\%$ confidence intervals are based on the full robust set.  Notice that the resampled correlated distribution creates nearly identical median values as obtained from the original robust set.  The median values for the remnant parameters in Table~\ref{tab: C220_bootstrap_value} will be used for each of the greedy fixed ${}^fC^+_{\ell{m}n}$ in subsequent iterations.  The error in these median values relative the simulation values is $\epsilon = 3.4 \times 10^{-5}$.

\begin{table}[ht!]
\renewcommand{\arraystretch}{1.5}
\begin{tabular}{ |c|c|c| } 
 \hline
 QNM  &$A^{+}_{\ell 20}$& $\phi^{+}_{\ell 20}$    \\ 
\hline 
    $C^{+}_{220}$&$0.9712^{+0.0009}_{-0.0011}$&$1.4829^{+0.0010}_{-0.0013}$\\
\hline
    $C^{+}_{320}$&$0.03782^{+0.00003}_{-0.00008}$&$-0.8279^{+0.0014}_{-0.0040}$\\
\hline
    $C^{+}_{420}$&$0.00198^{+0.00007}_{-0.00006}$&$-2.29^{+0.03}_{-0.05}$\\
\hline 
\multicolumn{3}{|c|}{Remnant Parameters} \\
\hline
$\delta$ & \multicolumn{2}{|c|}{$0.95199^{+0.00015}_{-0.00009}$} \\
\hline
$\chi_f$ & \multicolumn{2}{|c|}{$0.69207^{+0.00018}_{-0.00014}$} \\
\hline
\end{tabular}
\caption{Median values and $95\%$ confidence intervals for the amplitudes and phases of $C^+_{(2\sim4)20}$, and the remnant parameters.  Medians were computed by bootstrapping the $804$-element robust set created during the initialization step.  Conservative confidence intervals are taken directly form the robust set without bootstrapping.} \label{tab: C220_bootstrap_value}
\end{table}

\subsection{Nonlinear greedy iterations} \label{subsec: nonlinear_greedy_algorithm}
The initialization process outlined in Sec.~\ref{subsec: regular_nonlinear_fitting} is considered to be the zeroth iteration of the nonlinear fitting process.  The products of each iteration $i$ are
\begin{enumerate}
    \item A set of robust modes satisfying robustness criterion 3 for overtones with $n=i$;
    \item A robust set consisting of the union of sets from each model for which $\Delta_{NL}<0.01$ for the robust modes;
    \item A $500$-element correlated set of QNM values for each robust mode created by sampling the multivariate Gaussian copula distribution based on the robust set.
\end{enumerate}
For iteration $i+1$, a nonlinear fit is carried out for each element in the $500$-element set of robust modes from iteration $i$.  In each fit, the greedy fixed modes ${}^fC^\pm_{\ell{m}i}$ are taken from this set, while the greed fixed modes with $n<i$ are set to fixed values.  The values for ${}^fC^\pm_{\ell{m}n}$ with $n<i$ are set to the bootstrapped median values of their respective robust sets.

For iteration 0, the set of model results was taken from an extraction window within which the errors $\epsilon(t_i)$ in the estimated remnant parameters had attained a roughly consistent small value.  For iterations $i>0$, the set of model results was constructed by performing nonlinear greedy fits which maximized the partial overlap, and taking the results from the time $t_i$ which minimized $\epsilon$.  This produced a $500$-element set of fit values for each $C^\pm_{\ell{m}(i+1)}$ in each fit model, and this was used to determine which of these modes was robust within each model.

\subsubsection{First iteration results}
In the first iteration, the nonlinear fitting models fit the QNMs $\omega^+_{(2\sim4)2(0\sim4)}$ to signal modes $(2\sim4,2)$, where $N=1\sim4$.  The modes $\omega^+_{(2\sim4)20}$ were taken as greedy fixed modes, and the values of ${}^fC^+_{(2\sim4)20}$ were taken from the set of $500$ correlated modes randomly sampled from the six-dimensional Gaussian copula distribution constructed from the $804$-element robust set from iteration $0$.  Nonlinear greedy fitting was performed on each of the $500$ correlated greedy mode values for ${}^fC^+_{(2\sim4)20}$, maximizing the partial overlaps for each initial time $t_i$ in the range $0\le t_i\le35M$.  For each of the $500$ fits, a single set of values for the $n=1$ overtones was taken from the initial time $t_i$ corresponding to a local minimum\footnote{Spurious local minima at late times were ignored.} of the parameter error $\epsilon(t_i)$.  This created four sets (one for each model) of $500$ values for each $C^+_{(2\sim4)21}$.

\begin{figure*}[htbp]
    \centering
    \begin{subfigure}[t]{0.32\textwidth}
    \centering
    \includegraphics[width=\columnwidth]{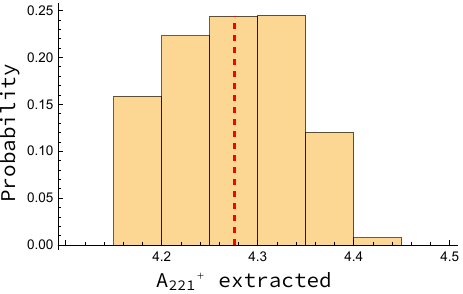}
    \end{subfigure}
    \hfill
    \begin{subfigure}[t]{0.32\textwidth}
    \centering
    \includegraphics[width=\columnwidth]{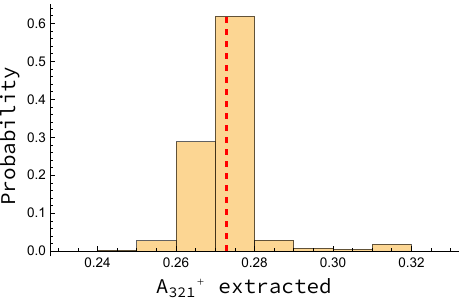}
    \end{subfigure}
    \hfill
    \begin{subfigure}[t]{0.32\textwidth}
    \centering
    \includegraphics[width=\columnwidth]{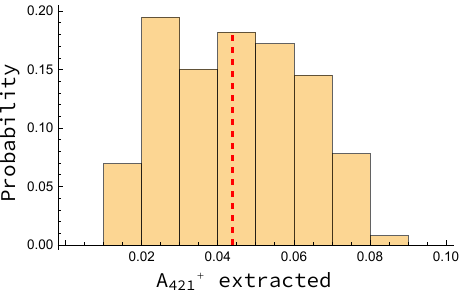}
    \end{subfigure}
    \caption{Probability distributions of $A^{+}_{(2\sim4)21}$ derived from the 1500 configurations of $C^+_{(2\sim4)21}$ generated in the $1_{\text{st}}$ iteration.}
    \label{fig:Hist_Amp221_321_421} 
\end{figure*}

Using just the models with $N=2\sim4$ we found that $C^+_{221}$ and $C^+_{321}$ satisfy robustness criterion 3 for all three models, but $C^+_{421}$ does not satisfy it for any model.  We created a $1500$-element robust set from the three sets of model data, from which we constructed bootstrapped medians for the amplitudes and phases, and for the uncertainties $\Delta_{NL}$.  The results are presented in Table~\ref{tab: C221_321_421_222_322_422_Median_CI_Nonlinear}.  The results include $95\%$ confidence intervals based on the data in the full robust set.  We also include the same construction for the nonrobust mode $C^+_{421}$.  Figure~\ref{fig:Hist_Amp221_321_421} presents the probability histograms of the amplitudes $A^+_{(2\sim4)21}$ from the same robust set.
\begin{table}[ht!]
\renewcommand{\arraystretch}{1.5}
\begin{tabular}{ |c|c|c|c|c| } 
 \hline
 Iteration &QNM  &$A^{+}_{\ell mn}$& $\phi^{+}_{\ell mn}$ & $\Delta_{NL}$  \\ [0.2cm]
 \hline
 \multirow{2}{*}{1} & $C^{+}_{221}$ & $4.28^{+0.11}_{-0.10}$ & $-0.661^{+0.018}_{-0.015}$ & 0.0028 \\[0.2cm]
    \cline{2-5}
 & $C^{+}_{321}$ & $0.273^{+0.027}_{-0.014}$ & $-2.92^{+0.13}_{-0.07}$ & 0.0038 \\[0.2cm]
    \cline{2-5}
 & $C^{+}_{421}$ & $0.044^{+0.033}_{-0.026}$ & $2.34^{+0.40}_{-0.35}$ & 0.057 \\[0.2cm]
 \hline
 \multirow{2}{*}{2} & $C^{+}_{222}$ & $12.8^{+9.0}_{-2.5}$ & $2.71^{+0.16}_{-0.50}$ & 0.034 \\[0.2cm]
   \cline{2-5}
 & $C^{+}_{322}$ & $0.8^{+1.0}_{-0.7}$ & $-0.2^{+3.0}_{-2.6}$ & 4.8 \\[0.2cm]
    \cline{2-5}
 & $C^{+}_{422}$ & $0.55^{+9.0}_{-0.34}$ & $0.0^{+2.1}_{-2.4}$ & 26 \\[0.2cm]
 \hline 
\end{tabular}
\caption{The medians, $95\%$ confidence intervals, and bootstrapping uncertainties $\Delta$ are given for the amplitudes $A^{+}_{\ell mn}$ and phases $\phi^{+}_{\ell mn}$ extracted from the first and second iteration of the nonlinear greedy method. Medians were computed by bootstrapping the combined dataset.  Conservative confidence intervals are taken directly form the dataset without bootstrapping.}\label{tab: C221_321_421_222_322_422_Median_CI_Nonlinear}
\end{table}

\subsubsection{Second iteration results}
In the second iteration, the nonlinear fitting models fit the QNMs $\omega^+_{(2\sim4)2(0\sim4)}$ to signal modes $(2\sim4,2)$, where $N=2\sim4$.  The modes $\omega^+_{(2\sim4)20}$ and $\omega^+_{(2\sim3)21}$ were taken as greedy fixed modes.  The values of ${}^fC^+_{(2\sim4)20}$ were fixed to the bootstrapped median values listed in Table~\ref{tab: C220_bootstrap_value}, while the values of ${}^fC^+_{(2\sim3)21}$ were taken from the set of $500$ correlated modes randomly sampled from the four-dimensional Gaussian copula distribution constructed from the $1500$-element robust set from iteration $1$.  Nonlinear greedy fitting was performed on each of the $500$ correlated greedy mode values for ${}^fC^+_{(2\sim3)21}$, maximizing the partial overlaps for each initial time $t_i$ in the range $0\le t_i\le35M$.  For each of the $500$ fits, a single set of values for the $n=2$ overtones was taken from the initial time $t_i$ corresponding to a local minimum of the parameter error $\epsilon(t_i)$.  This created three sets (one for each model) of $500$ values for each $C^+_{(2\sim4)22}$.

\begin{figure*}[htbp]
    \centering
    \begin{subfigure}[t]{0.32\textwidth}
    \centering
    \includegraphics[width=\columnwidth]{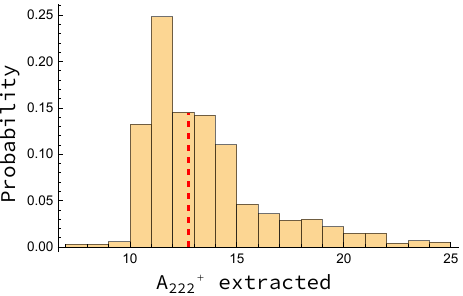}
    \end{subfigure}
    \hfill
    \begin{subfigure}[t]{0.32\textwidth}
    \centering
    \includegraphics[width=\columnwidth]{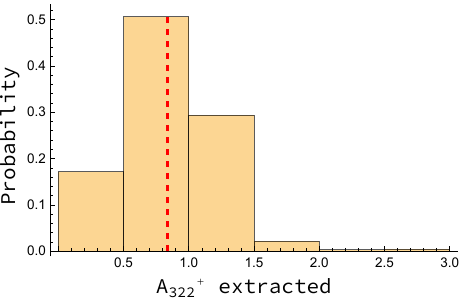}
    \end{subfigure}
    \hfill
    \begin{subfigure}[t]{0.32\textwidth}
    \centering
    \includegraphics[width=\columnwidth]{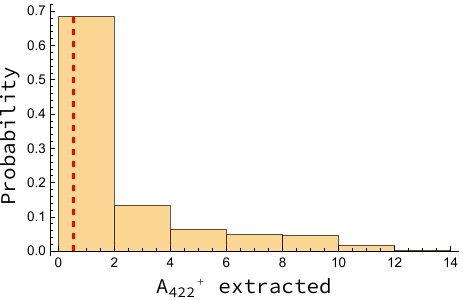}
    \end{subfigure}
    \caption{Probability distributions of $A^{+}_{(2\sim4)22}$ derived from the 1000 configurations of $C^+_{(2\sim4)22}$ generated in the second iteration.}
    \label{fig:Hist_Amp222_322_422} 
\end{figure*}

Using just the models with $N=3\sim4$ we found that none of the modes in any model satisfy robustness criterion 3.  For comparison, we created a $1000$-element robust set from the two sets of model data, from which we constructed bootstrapped medians for the amplitudes and phases, and for the uncertainties $\Delta_{NL}$.  The results are presented in Table~\ref{tab: C221_321_421_222_322_422_Median_CI_Nonlinear}.  The results include $95\%$ confidence intervals based on the data in the full robust set.  Figure~\ref{fig:Hist_Amp222_322_422} presents the probability histograms of the amplitudes $A^+_{(2\sim4)21}$ from the same nonrobust set.

\subsubsection{Terminating the iterations}
Because the second iteration found no robust modes, no further iterations of nonlinear greedy fitting were performed.  Interpretation of the nonlinear greedy fitting results are presented in the conclusions.

\section{Conclusion} \label{sec: conclusion}
In this work, we have investigated various aspects of the robustness of extracted QNM coefficients through ringdown fitting. In Secs.~\ref{subsec:robustness_criteria} and~\ref{subsec: compare_linear_models}, we introduced two robustness criteria for determining whether an extracted QNM can be considered robust in the linear fitting scenario. The first criterion evaluates the constancy of a QNM's expansion coefficients across an extraction window, with the fit start time varying across this window.  In Sec.~\ref{subsec: compare_linear_models}, we emphasized the importance of consistency among fitting models and formulated a second robustness criterion, which estimates the uncertainties of an extracted QNM coefficient across different models.  In Sec.~\ref{subsec: greedy_algorithm_description}, we provided a detailed description of a greedy algorithm for extracting QNM coefficients based on the eigenvalue method outlined in Ref.~\cite{Cook:2020otn}. This greedy approach applies to both linear and nonlinear fitting scenarios. By applying this method to the NR waveform SXS:BBH\_E\lowercase{xt}CCE:0305 in Sec.~\ref{subsec:robustness_and_the_greedy_algorithm}, we demonstrated that the robustness of overtone coefficients is enhanced by using greedy linear fitting, particularly for the $(2,2,2,+)$ overtone. The $(2,2,2,+)$ mode is the highest overtone that satisfies both robustness criteria. In Sec.~\ref{subsec: nonlinear_modes_210_210}, after we subtract the contribution of QNMs $(2\!\sim\!4,2,0,+)$ and $(2\!\sim\!3,2,1,+)$ from $h^{\text{NR}}_{42}$, we find indications of the possible presence of the quadratic QNM $(2,1,0,+)\times(2,1,0,+)$.

In Sec.~\ref{sec:nonlinear_fitting}, we extended the greedy approach to nonlinear fitting, employing it as an alternative method to generate a distribution of QNM coefficient values, and a third robustness criterion was applied to assess these sets of values. Overall, we found that the QNM coefficients estimated through nonlinear-greedy fitting were consistent with those from linear-greedy fitting. Notably, the conclusions regarding the robustness of the QNMs $(3,2,1,+)$ and $(2,2,2,+)$ differed slightly between the linear and nonlinear greedy approaches.  While the QNM $(3,2,1,+)$ was identified as only a tentative robust mode using linear-greedy fitting, this mode satisfied robustness criterion~3 across all three nonlinear fitting models. Furthermore, the estimated QNM coefficient $C^+_{321}$ from the nonlinear greedy fitting was consistent with the linear-greedy fitting result, providing additional evidence to support the validity of the estimated $C^+_{321}$ value. The second overtone $(2,2,2,+)$ exhibited larger uncertainties in the nonlinear fitting scenario, causing it to fail the robustness criterion. This outcome for overtone $(2,2,2,+)$ should not be surprising.  With each iteration of our nonlinear greedy procedure, the distribution of values being used is gaining breadth from prior iterations, inherently leading to increasing uncertainties.  In contrast, the uncertainty in the linear-greedy procedure does not suffer from a similar accumulation of uncertainty.

By going beyond the traditional criterion (our robustness criterion 1) related to the constancy of a QNM's expansion coefficients over a window in time, our various robustness criteria provide a more rigorous framework for identifying robust QNMs and faithfully estimating uncertainties in the extracted QNM coefficients.  As has been discussed already in the literature~\cite{Zhu:2024rej}, the first criterion can be viewed as providing a necessary, but not sufficient, condition for an extracted QNM coefficient to be considered robust. We have discussed and used additional robustness criteria.  However, we are not claiming that the specific criteria and thresholds we have used specify sharply defined necessary and sufficient conditions for an extracted QNM coefficient to be robust.

In this work, we have only applied our criteria and methods to a specific NR waveform: SXS:BBH\_E\lowercase{xt}CCE:0305, and several pure damped-sinusoid waveforms in Appendix~\ref{app:pure_damped_waveforms}. However, our approach can be easily applied to any NR waveforms, and we plan to apply our approach to entire catalogs of waveforms.  As mentioned in Sec.~\ref{subsec:simulation_data_used_for_fitting}, it is important for these waveforms to be mapped to the correct BMS frame to avoid unphysical mode mixing, and the super rest frame has been shown to be the correct choice of BMS frames for ringdown fitting.  By creating a catalog of robustly extracted QNM coefficients covering a large space of progenitor BBH systems, we hope to contribute to the understanding of how to parametrize the mapping of progenitor BBH systems to the excitation of specific QNMs in the remnant black hole.

During the preparation of this manuscript, we became aware of Ref.~\cite{Giesler:2024hcr} which explores the ability to extract both higher overtones and quadratic modes using an agnostic fitting process based on a variable projection algorithm.  In this work, the authors explored two different datasets, denoted as SXS:BBH:2423 and SXS:BBH:2420.  The methods and results presented in Ref.~\cite{Giesler:2024hcr} have notable similarities and differences to the work we present in this paper.  Both works explore the extraction of ringdown information from numerically generated binary black hole simulations based on general relativity, and ultimately extract gravitational QNMs of a remnant Kerr geometry.  An interesting difference is that in Ref.~\cite{Giesler:2024hcr}, the authors use an agnostic approach for determining which damped sinusoidal modes are present in the ringdown signal.  This approach removes the need to pre-select a set of linear and quadratic QNMs to include in a fitting model, although their method requires them to add some number of new agnostically determined damped sinusoids during each fitting iteration.  Another interesting difference is that a large number of overtones and quadratic modes are found in the $h_{22}$ signals.  We note  that one of the two signal explored in Ref.~\cite{Giesler:2024hcr} has a significantly larger remnant angular momenta than in the dataset we explored, while the other dataset has higher accuracy.  Larger angular momentum generally leads to longer damping times for the various modes, making it somewhat easier to extract more modes, while details of the progenitor system will also affect how many modes are significantly excited.  Also, it seems that in the approach used in Ref.~\cite{Giesler:2024hcr}, many of the higher tones and quadratic modes are shown to be generally consistent with the damping rates found in the set of agnostically extracted damped sinusoids, but few additional details are given.  No quantitative information is given as to the uncertainty in these modes, and many of the modes only display consistency over a relatively short time window.  While the focus of this paper has been on determining robust modes, the focus in Ref.~\cite{Giesler:2024hcr} was to show that when the ill-conditioned nature of fitting damped sinusoids is handled correctly, many modes can be successfully found even early in the ringdown. In Appendix~\ref{app:pure_damped_waveforms}, we briefly explore the ill-conditioned nature of fitting damped sinusoids. When fitting pure damped-sinusoidal waveforms, the robustness of the high overtones is sensitive to unmodeled modes (or noise). Our greedy approach outperforms the regular linear fitting in these ill-conditioned cases. 

An interesting similarity between our approach and that used in Ref.~\cite{Giesler:2024hcr} is that they are both based on methods which separate the underlying linear problem of fitting the complex mode amplitudes from the nonlinear fitting of the complex mode frequencies.  In the variable projection approach used in Ref.~\cite{Giesler:2024hcr}, each damped sinusoid included in the fitting model adds two linear degrees of freedom and two nonlinear degrees of freedom.  In our eigenvalue approach, we also have two linear degrees of freedom for each QNM or quadratic mode included in the fitting model.  Our approach then has two possibilities.  For linear fitting, we fix the mass and angular momentum of the remnant black hole and there is no nonlinear aspect to the fitting.  For nonlinear fitting, we have only two nonlinear degrees of freedom which are fixed by minimizing the mismatch.  This behavior may represent an advantage for the eigenvalue method, especially when multimode fitting is employed.  Our approach only has two nonlinear degrees of freedom regardless of how many modes are present in the fitting model and regardless of how many signal modes are used in the fit.  In Ref.~\cite{Giesler:2024hcr}, it seems that only a single mode, $h_{22}$, was fit and it is not clear how multimode fitting would be incorporated within the variable projection algorithm.

Another interesting similarity between the two approaches is that they both benefit from the use of a greedy approach, although they are implemented in different ways.  Both methods find it advantageous to start fitting at late fit-start times and work towards earlier fit-start times, fixing in some way the modes which have already been determined.  The primary difference is that our greedy approach fixes the expansion coefficient for each fixed mode, while the approach in Ref.~\cite{Giesler:2024hcr} is to fix the determined complex mode frequency, but leave the expansion coefficient free.  However, they emphasize that it is important that they also drop any rapidly damped modes from their fitting model for fit-start times where such modes should be negligible.  Where applicable\footnote{For the eigenvalue method, this only applies to nonlinear greedy fitting.}, both methods also use the remnant parameters obtained from the initial fits starting at late fit-start times.

It would be interesting to compare extracted complex mode amplitudes based on our methods and the methods used in Ref.~\cite{Giesler:2024hcr} when applied to the same datasets.  In particular, it would be interesting to better understand the level of uncertainty present in the higher overtones and quadratic modes.  Another interesting possibility is that the two methods are complementary.  The variable projection approach may provide a natural way to determine which set of QNMs and quadratic modes should be included in a fit model.  We leave this to future work.

\acknowledgments 
The authors would like to thank Mark Ho-Yeuk Cheung, Eliot Finch, Sizheng Ma, and Lorena Maga\~na Zertuche for their fruitful discussions. They also thank Keefe Mitman for clarifications on quadratic QNMs, and Saul Teukolsky for comments on the manuscript. Some computations were performed using the Wake Forest University (WFU) High Performance Computing Facility, a centrally managed computational resource available to WFU researchers including faculty, staff, students, and collaborators\cite{DEAC-Cluster}. This material is based upon work supported by the National Science Foundation under Grants No.~PHY-2407742, No.~PHY- 2207342, and No.~OAC-2209655, and by the Sherman Fairchild Foundation at Cornell; and under Grants No.~PHY-2309211, No.~PHY-2309231, and No.~OAC-2209656, and by the Sherman Fairchild Foundation at Caltech.  Any opinions, findings, and conclusions or recommendations expressed in this material are those of the author(s) and do not necessarily reflect the views of the National Science Foundation.

\appendix 
\section{DETAILS OF THE RESCALING} \label{app:details_rescaling}
The basic approach to rescaling during fitting has been explained in Sec.~\ref{subsec: rescaling}. In this appendix, we provide information regarding the practical implementation of rescaling and its associated benefits. All the examples are obtained using the eigenvalue method as described in Sec.~\ref{subsec:Waveform_Fitting}.

\begin{figure}[htb!]
\centering
\includegraphics[width=\columnwidth]{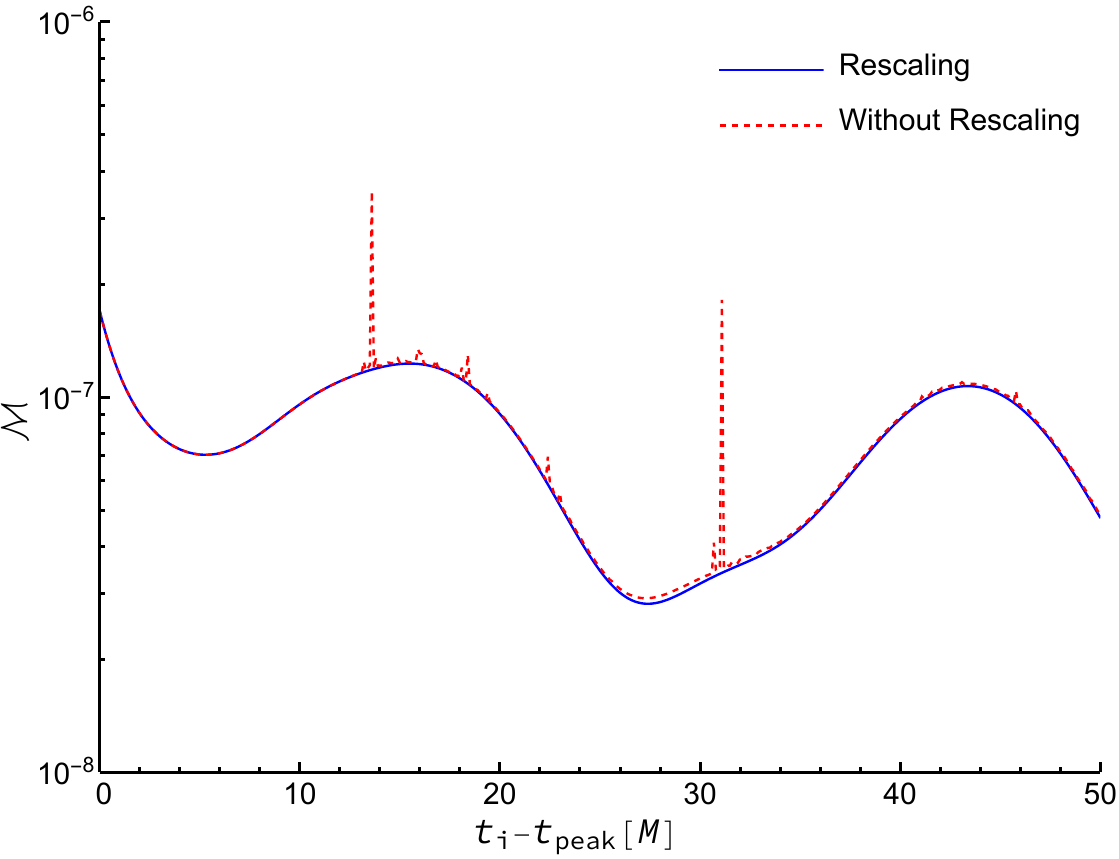}
\caption{The mismatches obtained from fitting QNMs $\omega^+_{(2\sim4)2(0\sim 7)}$ to signal modes $(2\sim4,2)$, plotted as functions of $t_i$. The solid blue line is obtained from linear fitting with rescaling. The dashed red line is obtained from linear fitting without rescaling. Machine precision was used during the singular value decompositions.}
\label{fig:C42QNM42N7MismatchRescaling}
\end{figure}

After applying singular value decomposition, the $n\times n$ Hermitian matrix $\mathbb B$ defined in Eq.~\eqref{eqn:Definition_B} is decomposed as $U\Sigma U^{\dagger}$, where $U$ is $n\times n$ complex unitary matrix, and $\Sigma$ is a $n\times n$ diagonal matrix with non-negative real singular values $\sigma_i$ on the diagonal.  The singular value decomposition result for the matrix $\mathbb B$ is special because the Hermitian matrix $\mathbb B$ for our case 
is positive definite.

The inverse matrix $\mathbb B^{-1}$ needed in Eq.~\eqref{eqn:least-squares_coefs} is obtained using $U \Sigma^{-1} U^{\dagger}$. The condition number of a matrix is defined as the ratio of the largest singular value to the smallest singular value. The matrix is ill-conditioned if the condition number is too large.  When a large set of higher overtones and subdominant modes are included in the fitting model, the matrix $\mathbb{B}$ can be ill-conditioned, which will cause noticeable roundoff error during the SVD. 

The first benefit of rescaling is to decrease the noise in the fitting results without requiring extra precision for the SVD.  Using extra precision when performing the SVD leads to a much longer computing times.  As an example, we fit QNMs $\omega^+_{(2\sim4)2(0\sim 7)}$ to signal modes $(2\sim4,2)$ and explore the results in several figures. Figure~\ref{fig:C42QNM42N7MismatchRescaling}, displays the mismatch as computed with and without rescaling.  We obtained both results using machine precision in Mathematica when computing the SVD.  Without the use of rescaling, achiving a smooth mismatch requires a SVD precision of about 20 decimal digits, which results in a factor of 60 increase in running time.

\begin{figure}[htb!]
\centering
\includegraphics[width=\columnwidth]{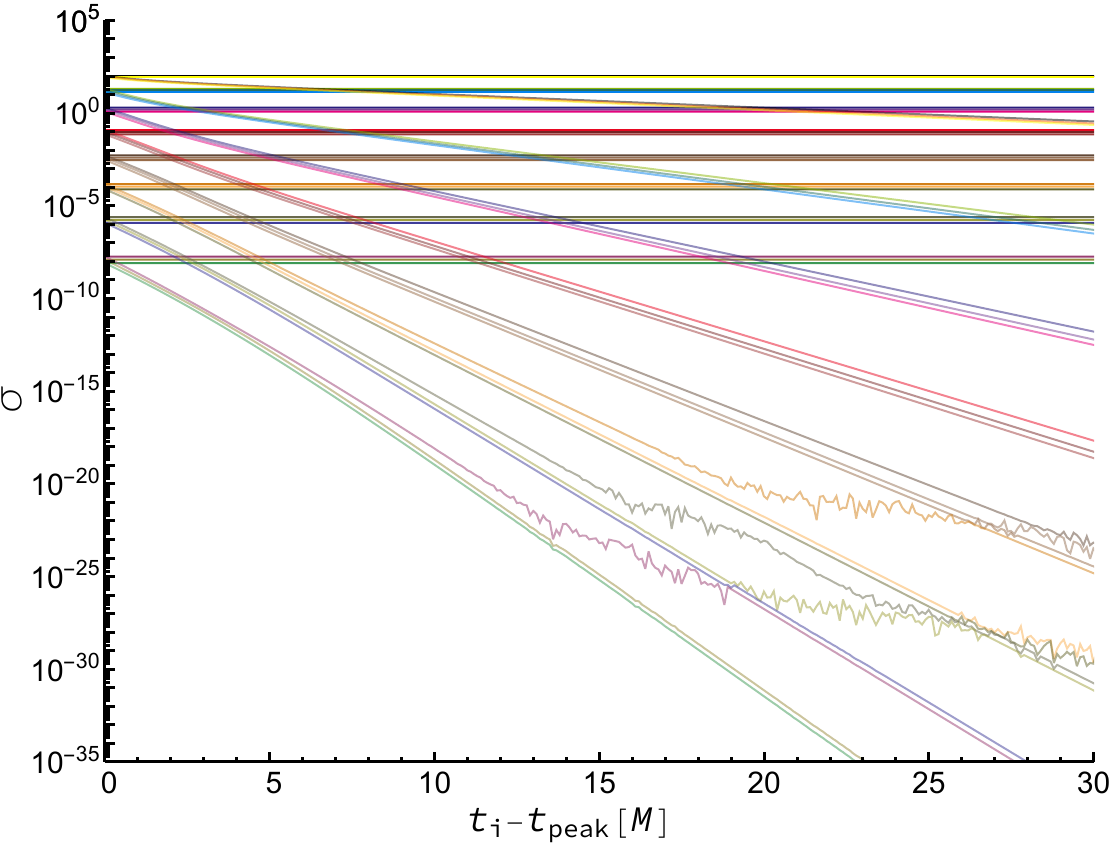}
\caption{The singular values of $\mathbb{B}$ as computed both with and without rescaling, plotted as functions of $t_i$. The QNM set $\omega^+_{(2\sim4)2(0\sim 7)}$ is used to construct the matrix $\mathbb{B}$. The horizontal lines are the singular values obtained with rescaling. The exponentially damping lines are obtained without rescaling. Machine precision was used during the singular value decompositions.}
\label{fig:C42QNM42N7SingularValues}
\end{figure}

In Fig.~\ref{fig:C42QNM42N7SingularValues}, we display the singular values obtained from applying SVD to $\mathbb{B}$ both with and without rescaling, and plot the results as functions of the fit-start time $t_i$. The singular values obtained without rescaling show dramatic exponential decay to a very small value, and there is clearly evidence of roundoff errors in the singular values computed without rescaling. In contrast, the singular values obtained with rescaling are largely constant throughout $t_i$. This explains why the noise in the mismatch is reduced by incorporating the rescaling process.

\begin{figure}[htb!]
\centering
\includegraphics[width=\columnwidth]{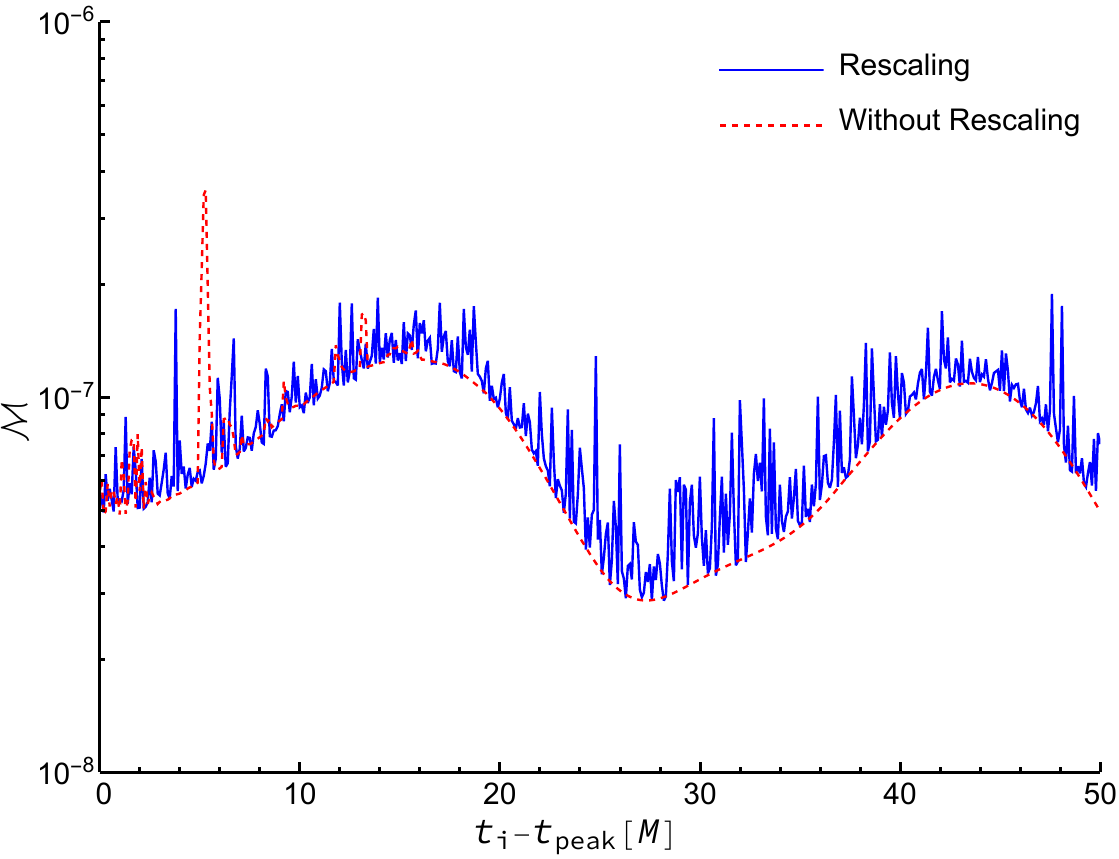}
\caption{The mismatches obtained from fitting QNMs $\omega^{\pm}_{(2\sim4)2(0\sim 6)}$ to signal modes $(2\sim4,2)$, plotted as functions of $t_i$. The solid blue line is obtained from linear fitting with rescaling. The dashed red line is obtained from linear fitting without rescaling. Machine precision was used during the singular value decompositions.}
\label{fig:C42QNM42N6PMMismatchRescaling.pdf}
\end{figure}

\begin{figure}[htb!]
\centering
\includegraphics[width=\columnwidth]{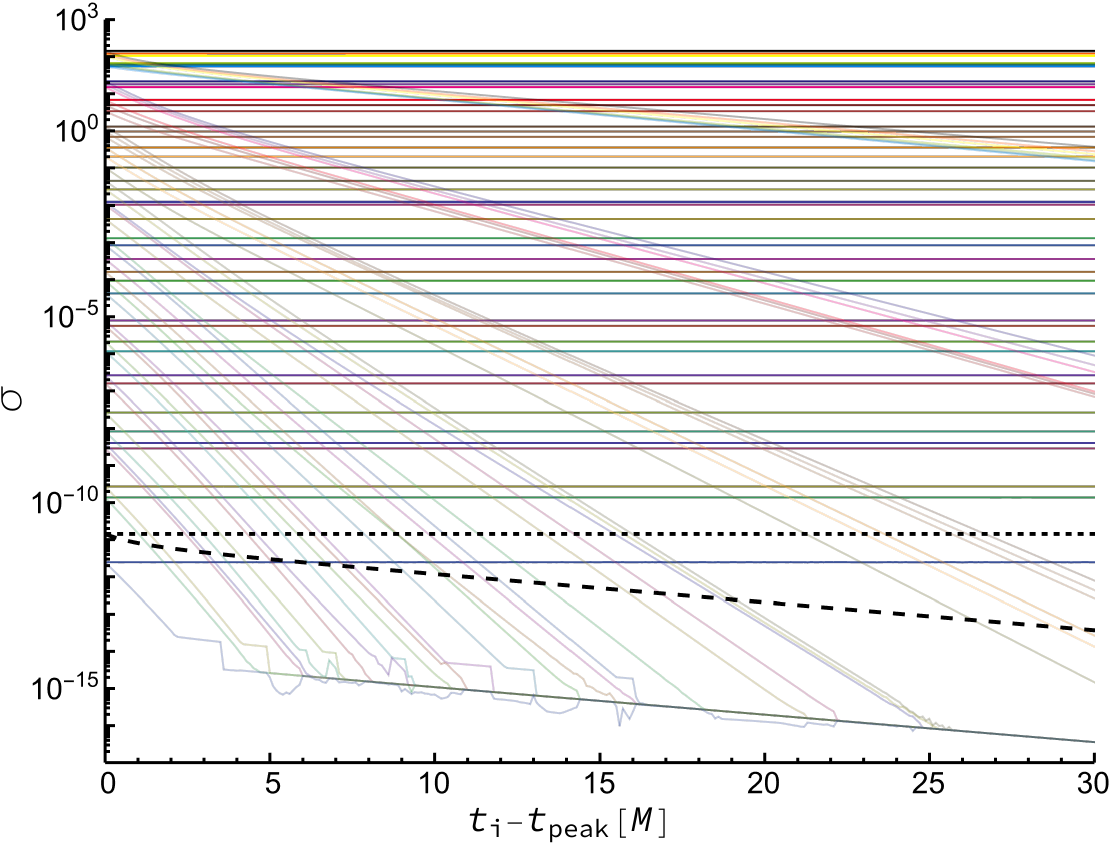}
\caption{The singular values of $\mathbb{B}$ as computed both with and without rescaling, plotted as functions of $t_i$. The QNM set $\omega^\pm_{(2\sim4)2(0\sim 7)}$ is used to construct the matrix $\mathbb{B}$. The horizontal lines are the singular values obtained with rescaling. The exponentially damping lines are obtained without rescaling. Machine precision was used during the singular value decompositions.  The dashed lines show the cutoffs for an SVD tolerance of $10^{-13}$ used used when computing $\mathbb{B}^{-1}$. The short dashed line is for the case with rescaling, while the longer dashed line is for the case without rescaling. }
\label{fig:C42QNM42N6PMSingularValues}
\end{figure}

When we extend the fitting model to include overtones of the retrograde QNMs, there are cases where significant noise is present in the mismatch, even with rescaling. In Fig.~\ref{fig:C42QNM42N6PMMismatchRescaling.pdf}, we consider the example of fitting QNMs $\omega^{\pm}_{(2\sim4)2(0\sim 6)}$ to signal modes $(2\sim4,2)$.  Note that such a fitting models are unlikely to be used in a real application since so many subdominant retrograde overtones are unlikely to be present. Often, rescaling alone can deal very well with noise in the fittings.  However, this example illustrates how we can deal with excessive noise by setting a SVD tolerance. In Fig.~\ref{fig:C42QNM42N6PMMismatchRescaling.pdf}, we see that the mismatch from fitting with and without rescaling both have noise, and the mismatch with rescaling actually shows an increased level of noise. In Fig.~\ref{fig:C42QNM42N6PMSingularValues}, we plot the singular values from the SVD of $\mathbb{B}$ both with and without rescaling as functions of $t_i$. The exponentially decaying singular values obtained without rescaling show obvious discontinuities throughout $t_i$, while the singular values obtained with rescaling appear smooth.  We can improve the fitting results for both cases by setting a tolerance level for the singular values which will affect the computation of the pseudoinverse $\mathbb{B}^{-1}$.

\begin{figure}[htb!]
\centering
\includegraphics[width=\columnwidth]{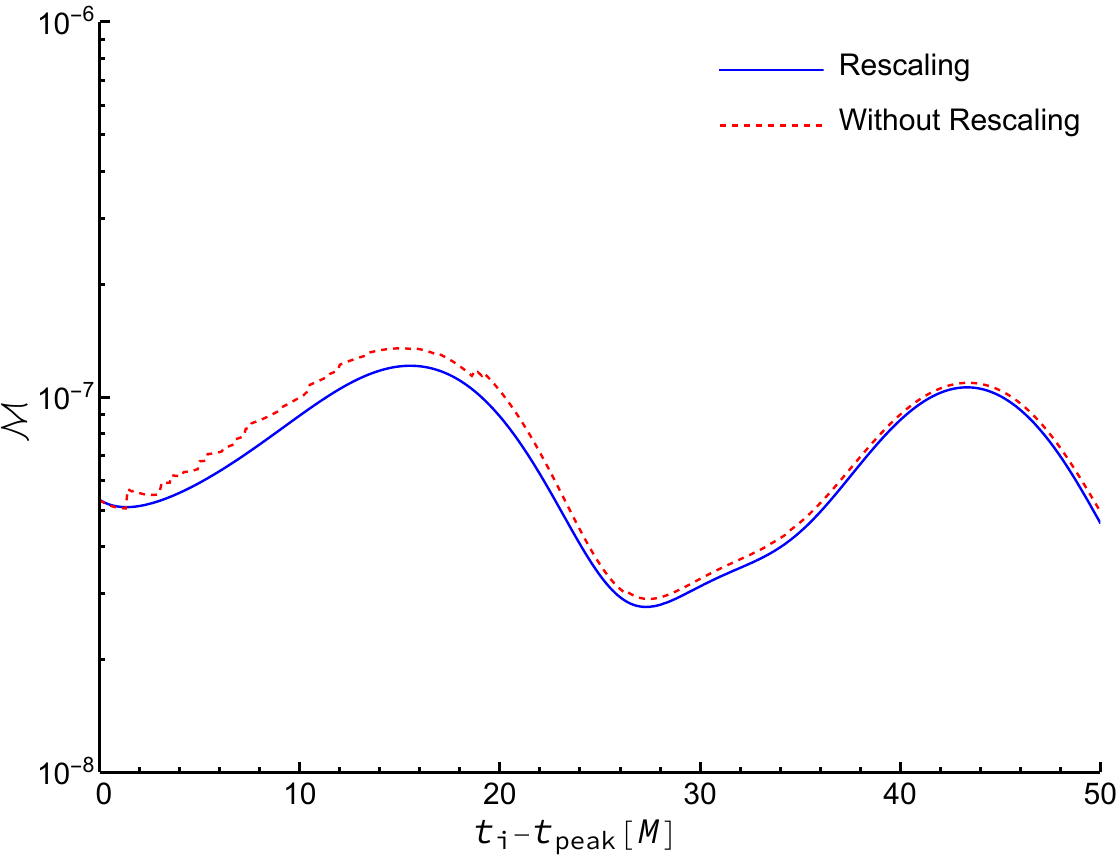}
\caption{The mismatchs obtained from fitting QNMs $\omega^{\pm}_{(2\sim4)2(0\sim 6)}$ to signal modes $(2\sim4,2)$ computed with an SVD tolerance of $10^{-13}$, plotted as functions of $t_i$. The solid blue line is obtained from linear fitting with rescaling. The dashed red line is obtained from linear fitting without rescaling. Machine precision was used during the singular value decompositions.}
\label{fig:C42QNM42N6PMMismatchRescalingTol13.pdf}
\end{figure}

When computing the pseudoinverse $\mathbb{B}^{-1}$, a SVD tolerance determines when the inverse singular value $1/\sigma_i$ in $\Sigma^{-1}$ will be set to $0$.  If the ratio of a particular singular value $\sigma_i$ to the largest singular value is smaller than the tolerance number, then we make the replacement $1/\sigma_i\to0$. In Fig.~\ref{fig:C42QNM42N6PMMismatchRescalingTol13.pdf}, we set the SVD tolerance to be $10^{-13}$ and recompute the mismatch with all other settings the same as those used in creating Fig.~\ref{fig:C42QNM42N6PMMismatchRescaling.pdf}. The inverse of any singular values below the dashed lines in Fig.~\ref{fig:C42QNM42N6PMSingularValues} are set to $0$ when we compute $\mathbb{B}^{-1}$.  The noise in the mismatch is removed for both cases. However, for the case without rescaling, we can see that there are discontinuities remaining in the mismatch. This is because the singular values without rescaling exponentially decay as $t_i$ increases, and the number of $1/\sigma_i$ that are set to $0$ continually changes. In contrast, only the one smallest singular value in the case with rescaling has its inverse set to $0$.  This happens at every $t_i$, which yields a smooth mismatch curve. Again, the benefit of using SVD tolerance to obtain smooth fitting results instead of increasing SVD precision is that it avoids a dramatic increase in computation time. 

\section{TESTING WITH PURE DAMPED-SINUSOID WAVEFORMS} \label{app:pure_damped_waveforms}
In this paper, we have been fitting to a waveform simulated in full nonlinear GR. In this appendix, we apply our linear fitting approach to test waveforms that are a linear combination of pure damped sinusoids. These waveforms exactly follow linear perturbation theory, with no nonlinear effects, no power-law tails, and with numerical noise only arising from finite precision numerics. The lessons we learned from this type of fitting can help us interpret the fitting results obtained from full NR waveforms. Furthermore, we show the advantages of the greedy approach in improving the robustness of the extracted overtones and subdominant modes in this controlled setting. 

One might expect that the injected QNM coefficients of the test waveform will all be extracted robustly with good precision if we choose a fitting model exactly matching the set of injected QNMs. However, there are still limitations in fitting the high overtones and subdominant modes in this ideal scenario. 
We first considered a test waveform consisting of the injected QNMs $\omega^+_{(2\sim4)2(0\sim4)}$.\footnote{The amplitudes and phases for the test waveform are chosen based on the results from Sec.~\ref{subsubsec: the_importance_of_multimode_fitting}.  In particular, we use the model fit median values from Table~\ref{tab: C220_Summary_Table} for the robustly fit modes, and representative extracted values for the nonrobustly fit modes.} Using simple linear fitting, and allowing for mode mixing for $\ell=2\sim4$, we found that the QNMs $(2,2,0\!\!\sim\!\!4,+)$, $(3,2,0\!\!\sim\!\!1,+)$, and $(4,2,0,+)$ could be fitted robustly and extracted with good agreements to the injected values.  After the greedy algorithm was applied, the QNM $(4,2,1,+)$ became marginally robust. Its $\Delta_\text{min}$ decreased to $0.0083$ after we fixed $C^+_{(2\sim4)20}$ for fitting model $\omega^+_{(2\sim4)2(0\sim2)}$. However, that was the only fitting model satisfying our robustness criterion $1$. In comparison, the $\Delta_\text{min}$ of $C^+_{421}$ obtained from same regular fitting was $0.0335$.  
%The plot is not significant so might not need to be included here. 

\begin{figure}[htb!]
\centering
\includegraphics[width=\columnwidth]{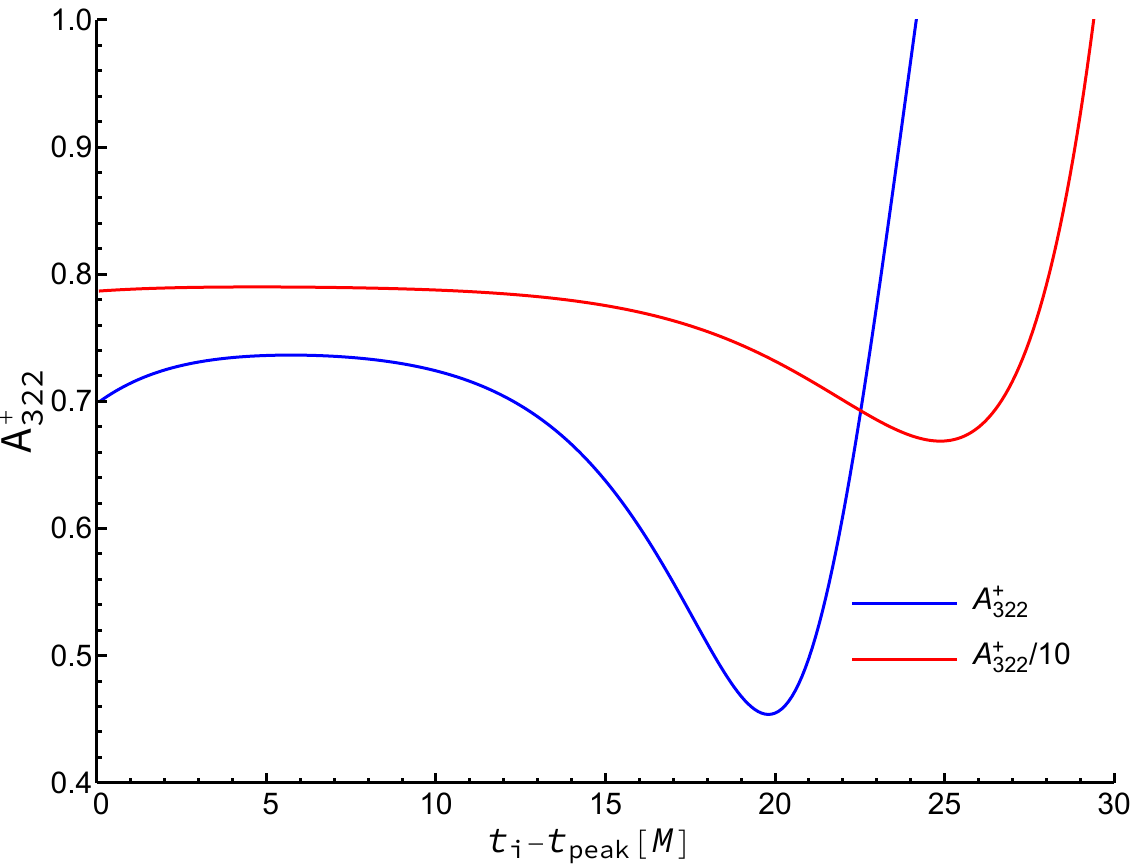}
\caption{The amplitude $A^+_{322}$ is plotted as a function of the initial fitting time $t_i$. The blue line is obtained from fitting the model waveform incorporating $A^+_{322}=0.8$, while the red line is obtained from fitting the model waveform with $A^+_{322}=8$. For ease of comparison, the value of $A^+_{322}$ for the red line has been divided by a factor of 10. 
For both case, the fitting model is fitting QNMs $\omega^+_{(2\sim4)2(0\sim4)}$ to signal modes $(2\sim4,2)$ with $C^+_{(2\sim4)20}$ and $C^+_{(2\sim3)21}$ fixed to values obtained from prior greedy iterations. }
\label{fig:CompareA322Appendix}
\end{figure}

The subdominant overtones, such as $(3,2,2,+)$ and $(4,2,1,+)$, have fast damping rates and have small amplitudes in the model we considered. These two facts are important reasons that the subdominant higher overtones are hard to recover even in this model waveform. In a further test, we increased the injected amplitudes of $(3,2,2,+)$ from $0.8$ to $8$ and kept all the other QNMs the same. We found that the QNM $(3,2,2,+)$ could then be extracted more robustly and with higher accuracy for the larger amplitude case, as shown in Fig.~\ref{fig:CompareA322Appendix}. 
%Attach one figure to show this effects here. 
A similar pattern was found in other subdominant overtones when their amplitudes were increased. 

\begin{figure}[htb!]
\centering
\includegraphics[width=\columnwidth]{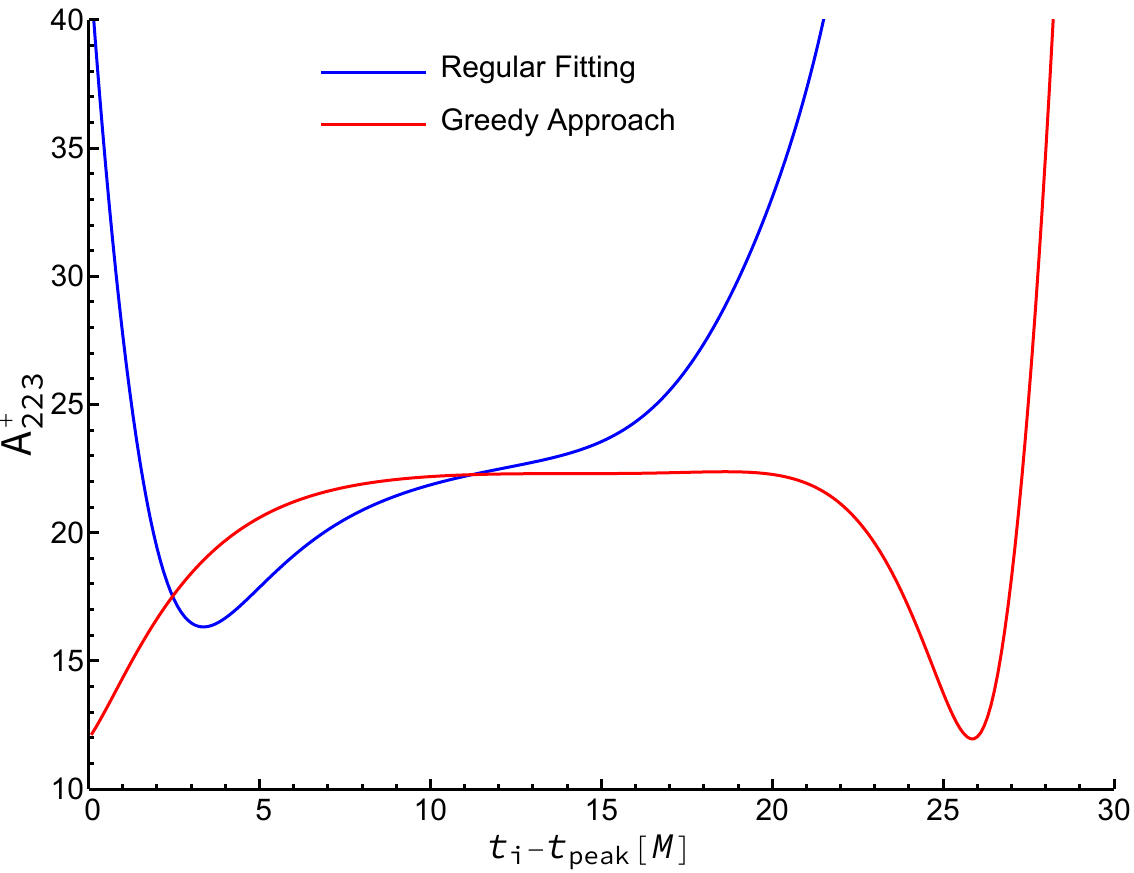}
\caption{The amplitude $A^+_{223}$ is plotted as a function of the initial fitting time $t_i$. In this case, the QNM $(2,2,5,+)$ has been added to the model waveform as an unmodeled source of noise.  The blue line is obtained from regular fitting, while the red line is obtained from greedy fitting with $C^+_{(2\sim4)20}$, $C^+_{(2\sim3)21}$, and $C^+_{222}$ fixed to their values obtained from prior greedy iterations.
For both cases, the fitting model is fitting QNMs $\omega^+_{(2\sim4)2(0\sim4)}$ to signal modes $(2\sim4,2)$. }
\label{fig:CompareA223GreedyAppendix}
\end{figure}

In a final test, we introduced an additional overtone to mimic the noise from higher overtones in the early stage of ringdown. In this test, the additional overtone $(2,2,5,+)$ was injected in the model waveform with a large amplitude of $A^+_{225}=28.93$ but without including it in the fitting models. 
After this noise source was introduced, the overtones $(2,2,3\!\!\sim\!\!4,+)$ were no longer robust using regular linear fitting. In contrast, the coefficient of $(2,2,3,+)$ could be extracted robustly, and close to the injected amplitude $A^+_{223}=22.59$, by applying the greedy approach. A direct comparison for the amplitude $A^+_{223}$ obtained from regular and greedy fitting is shown in Fig.~\ref{fig:CompareA223GreedyAppendix}.
%Use a plot to prove this statement 

In conclusion, these tests with model waveforms suggest that the subdominant modes are hard to extract robustly in a real signal not only due to the noise and unconsidered effects in the ringdown but also because of their relatively small amplitudes within the waveform. Furthermore, robust extraction of higher overtones are sensitive to noise, such as unconsidered overtones in the waveform, but the greedy algorithm 
can help lower the uncertainties in extracting these modes, increasing our chances of extracting more robust QNMs from real ringdown signals.

\end{document}